\documentclass[showpacs,amsmath,amssymb,twocolumn,pra,superscriptaddress]{revtex4-1}

\usepackage[dvips]{graphicx} 
\usepackage{amsfonts}
\usepackage{amssymb}
\usepackage{amscd}
\usepackage{amsmath}    
\usepackage{enumerate}
\usepackage{epsfig}
\usepackage{subfigure}
\usepackage{xcolor}
\usepackage{amsthm}
\usepackage{framed}
\usepackage{multirow}

\begin{document}
\preprint{APS/123-QED}
\title{Clauser-Horne Bell test with imperfect random inputs}

\date{\today}
\author{Xiao Yuan}
\author{Qi Zhao}
\author{Xiongfeng Ma}
\affiliation{Center for Quantum Information, Institute for Interdisciplinary Information Sciences, Tsinghua University, Beijing 100084, China}

\begin{abstract}
Bell test is one of the most important tools in quantum information science. On the one hand, it enables fundamental test for the physics laws of nature, and on the other hand, it can be also applied in varieties of device independent tasks such as quantum key distribution and random number generation.
In practice, loopholes existing in experimental demonstrations of Bell tests may affect the validity of the conclusions.
In this work, we focus on the randomness (freewill) loophole and investigate the randomness requirement in a well-known Bell test, the Clauser-Horne test, under various conditions.
With partially random inputs, we explicitly bound the Bell value for all local hidden variable models by optimizing the classical strategy.
Our result thus puts input randomness requirement on the Clauser-Horne test under varieties of practical scenarios.
The employed analysis technique can be generalized to other Bell's inequalities.
\end{abstract}

\maketitle

\section{Introduction}
Since the inception of quantum mechanics, whether the law of nature is deterministic or truly random has been long debated. During this debate, Einstein, Podolsky, and Rosen (EPR) proposed a paradox  \cite{Einstein35} that eventually leaded to a counterintuitive phenomenon --- quantum nonlocality. Later, Bell put the EPR paradox in an experimentally testable framework, known as Bell test \cite{bell}.
In the bipartite scenario, a Bell test involves two remotely separated parties, Alice and Bob, who receive random inputs $x$ and $y$ and produce outputs $a$ and $b$, respectively. Based on the probability distribution $\tilde{p}_{AB}(a,b|x,y)$ of the outputs conditioned on the inputs, Bell's inequality can be defined by a linear combination of $\tilde{p}_{AB}(a,b|x,y)$ according to
\begin{equation}\label{eq:Bell}
J = \sum_{a,b,x,y} \beta_{a,b}^{x,y}\tilde{p}_{AB}(a,b|x,y) \leq J_C,
\end{equation}
where $J_C$ is a bound for all  local hidden variable models (LHVMs), meaning that, any LHVM cannot violate any Bell's inequality.

The essence of a Bell test is that the corresponding inequality can be violated with certain quantum settings, thus indicating the incompleteness of LHVMs.  That is, when Alice and Bob share some appropriate quantum state $\rho_{AB}$ and perform proper quantum measurements $M_x^a$ and $M_y^b$, the probability distribution $\tilde{p}_{AB}(a,b|x,y) = \mathrm{Tr}[\rho_{AB}M_x^a\otimes M_y^b]$ can achieve a higher Bell value $J > J_C$, thus violate the Bell's inequality defined in Eq.~\eqref{eq:Bell}.

The violation of Bell's inequality not only acts as a test for fundamental laws of physics, but has varieties of  applications in modern quantum information tasks. For instance, observing violations of Bell's inequalities can be used in device independent tasks, such as quantum key distribution \cite{Mayers98, acin06,masanes2011secure,Vazirani14}, randomness amplification \cite{colbeck2012free, gallego2013full, Dhara14} and generation \cite{Colbeck11, Fehr13, Pironio13, vazirani2012certifiable}, entanglement quantification \cite{Moroder13}, and dimension witness \cite{Brunner08}.

Therefore, experimental realization of Bell tests has both theoretical and practical significance. Amongst most of the experiment demonstrations, the well-known Clauser-Horne-Shimony-Holt (CHSH) inequality \cite{CHSH} is generally favored for Bell tests within bipartite owing for its simple and symmetric definition. Since the first experiment in the  early 1980s \cite{Aspect1982PhysRevLett.49.91}, violation of the CHSH inequality have been realized in varieties of experiment systems, including optic \cite{Christensen13,giustina2013bell},  superconducting \cite{ansmann2009violation}, ionic \cite{rowe2001experimental}, and atomic \cite{hofmann2012heralded} systems. However, these experiments suffer from a few technical and inherent loopholes \cite{Larsson14, Kofler14},  which might invalidate the conclusions.

There are three main inherent loopholes. First, the locality loophole refers to the scenario where testing devices are not far away enough from each other so that hidden signaling might be possible. With hidden signaling between devices in Bell tests, local hidden variables become global ones and hence the Bell's inequalities can be violated. This loophole could be closed by separating the  parties sufficiently apart with regard to the synchronization precision of different measurements in the tests.  That is, the operations on each party of the Bell tests should be spacelikely separated. In experiments, the locality loophole has been closed with entangled photons \cite{Weihs98} and is shown to be possible to close with atomic systems \cite{hofmann2012heralded}.

Second, the detection efficiency loophole stems from the inefficiency of detection systems. The statistics of the undetected \cite{PhysRevD.2.1418} could be different from the detected ones. For instance, it is shown that maximally entangled states cannot violate the CHSH test with detection efficiency lower than $84\%$ \cite{Garg87}. When considering non-maximally entangled states and the contribution of loss, the Clauser-Horne (CH) \cite{CH74} or Eberhard's \cite{Eberhard93} inequality has an advantage over the CHSH inequality, since it can tolerate lower detection efficiencies. Nevertheless, it is proved \cite{Massar03, Wilms08} that $2/3$ is the minimum requirement for detection efficiency when considering bipartite Bell tests with bit inputs and bit outputs. The counter-measure of the efficiency loophole is to improve the efficiency of the instruments to satisfy a critical requirement, where even using imperfect instruments, the result of a Bell test with a quantum strategy cannot be achieved by any LHVM strategy. Closing the efficiency loophole has been experimentally demonstrated \cite{rowe2001experimental, Christensen13,giustina2013bell}. Although closing both of the locality loophole and the efficiency loophole simultaneously is yet to be demonstrated, such test is practically feasible with current technology and could be realized in the near future.

Third, the randomness (freewill) loophole refers to the underlying assumption in Bell tests that different measurement settings can be chosen randomly (freely). Generally, a Bell test requires the inputs of each party to be fully random in order to avoid information leakage between different parties. If there is a local hidden variable that shares information about the random inputs, where in the worst scenario, the inputs are all predetermined such that each party knows exactly the inputs of the other party, it is possible to violate Bell inequalities just with LHVMs. Since one can always argue that there might exist a powerful creator who determines everything including all the Bell test experiments, this loophole is widely believed to be impossible to close perfectly. In this case, as we cannot prove or disprove the existence truly input randomness, the assumption of freewill is indispensable in general Bell tests.

Yet, it is still meaningful to discuss the randomness requirement \footnote{The imperfect input randomness requirement is sometimes called measurement dependence in literature.} of Bell tests in a practical scenario. In this case, we suppose that the randomness generation devices are partially controlled by an adversary Eve, who thus has partial knowledge of Alice's and Bob's inputs. Then she can make use of the information about the inputs to fake violations of Bell's inequalities \cite{Hall10} and thus lead to the device independent tasks insecure. Therefore, it is interesting to see how much of randomness needed for a Bell test in order to ensure the correctness of the conclusion. This is especially meaningful when considering a loophole free Bell test \cite{Larsson14, Kofler14} and its applications to practical tasks in the presence of an eavesdropper.

In experiment, as the CH test with non-maximally entangled states can tolerate lower detection efficiencies, it is generally considered as a promising candidate \cite{giustina2013bell, Christensen13} for implementing a loophole-free Bell experiment. Therefore, in this paper, we mainly focus on the CH test \cite{CH74} and investigate the randomness requirements under different circumstances. For the CH inequality, we show explicitly the optimal strategy with LHVM to maximize the Bell value. Our result thus put requirements on experimental realization of a loophole-free Bell test and other applications of device-independent tasks. Note that our analysis method can be easily generalized to other Bell's inequalities.


The rest of the paper is organized as follows. In Section \ref{Sec:Randomness}, we review previous works and introduce the quantification of the requirement of input randomness. In Section \ref{Sec:CH}, we analyze the randomness requirement for the CH test under different conditions. In Section \ref{Sec:conclusion}, we conclude our result and discuss the applications.

\section{Randomness Requirement}\label{Sec:Randomness}
\begin{figure}[thb]
\centering
\resizebox{3.5cm}{!}{\includegraphics[scale=1]{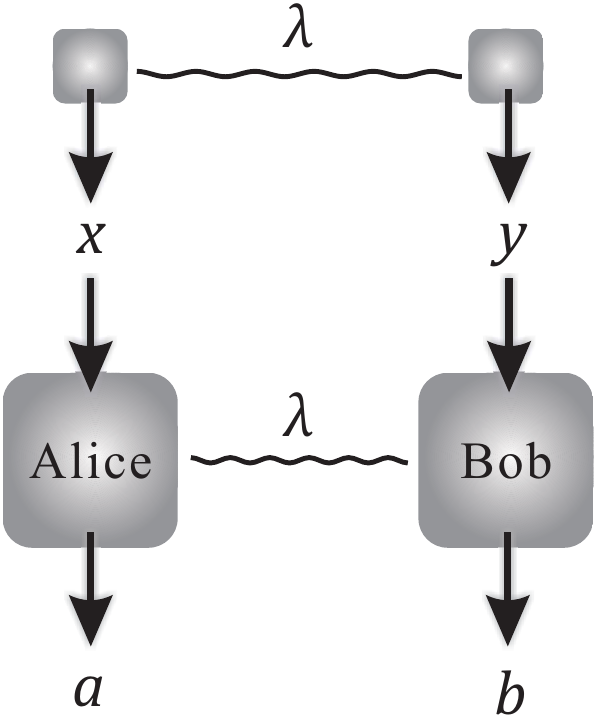}}
\caption{Bell tests in a bipartite scenario. In general, the inputs depend on some local hidden variable $\lambda$. The local hidden variables that control the inputs and the devices may be different. While, we can still denote these two local hidden variables with a single one denoted as $\lambda$.}\label{Fig:BellTest}
\end{figure}

In this work, we consider Bell's inequalities with input settings not chosen fully randomly. That is, the inputs $x$ and $y$ depend on some local hidden variable, denoted as $\lambda$, as shown in Fig.~\ref{Fig:BellTest}. In one extreme case, when  $\lambda$ deterministically decides the inputs, any adversary who access  $\lambda$ can fake violations of arbitrary Bell inequalities even with LHVMs. In the other extreme case, when the inputs are independent of $\lambda$, Bell's inequalities cannot be violated with any LHVMs. Then we can see that the correctness of the conclusion of Bell tests relies on how random the inputs are.

Here, we first quantify randomness by the dependence of the inputs conditioned on $\lambda$. Suppose the inputs $x$ and $y$ are chosen according to \emph{a priori} probability $p(x,y|\lambda)$, the input randomness can be quantified by its upper and lower bounds,
\begin{equation}\label{eq:randomness}
\begin{aligned}
P = \max_{x,y,\lambda}p(x,y|\lambda),\\
Q = \min_{x,y,\lambda}p(x,y|\lambda).
\end{aligned}
\end{equation}
As an example, for the CH test, where the inputs are binary, the upper and lower bounds are in the range of $[1/4, 1]$ and $[0, 1/4]$, respectively. Focusing on the upper bound $P$, when it equals $1$, it represent the case that the local hidden variable $\lambda$  deterministically decides at least one input. When $P = 1/4$, this corresponds to the case that the inputs are fully random. Similarly, we can see how the lower bound $Q$ characterizes the input randomness. In many of previous works \cite{Hall10,Hall11,Koh12,Pope13, Thinh13, Yuan15}, only the upper bound $P$ is considered. It is recently Note in Ref.~\cite{putz14} that the lower bound $Q$ also plays an important role in analysis. We thus consider both the upper and lower bounds as quantifications of the input randomness.

With binary inputs, we can consider a symmetric case where $P = 1/4 + \delta$ and $Q = 1/4 - \delta$. In other words, we can quantify the input randomness by its deviation from  a unform distribution, quantified by $\delta$,
\begin{equation}\label{Eq:deltasource}
  \delta = \max_{x,y,\lambda} \left|p(x,y|\lambda) - \frac{1}{4}\right|.
\end{equation}
Note that all our following results apply for asymmetric cases (with arbitrary $P$ and $Q$) as well.

When the input settings are determined by $p(x,y|\lambda)$, the observed probability $\tilde{p}_{AB}(a,b|x,y)$ of outputs conditioned on inputs is given by
\begin{equation}\label{Eq:pabxy}
   \tilde{p}_{AB}(a,b|x,y) = \frac{\sum_\lambda \tilde{p}_{AB}(a,b|x,y,\lambda)p(x,y|\lambda)q(\lambda)}{p(x,y)},
\end{equation}
where $q(\lambda)$ is the priori probability of $\lambda$, $p(x,y) = \sum_\lambda p(x,y|\lambda)q(\lambda)$ is the averaged probability of choosing $x$ and $y$, and $\tilde{p}_{AB}(a,b|x,y,\lambda)$ is the strategy of Alice and Bob conditioned on $\lambda$. Then, the Bell's inequality defined in Eq.~\eqref{eq:Bell} should be rephrased by
\begin{equation}\label{Eq:BellFree}
\begin{aligned}
J &= \sum_{x,y} \frac{1}{p(x,y)}\sum_\lambda\sum_{a,b} \beta_{a,b}^{x,y}\tilde{p}_{AB}(a,b|x,y,\lambda)p(x,y|\lambda)q(\lambda) \\
  &\leq J_C.
\end{aligned}
\end{equation}

In this work, we are interested in how LHVMs can fake a violation of Bell's inequality with imperfect input randomness. Thus, we can also set the strategy $\tilde{p}_{AB}(a,b|x,y,\lambda)$ of deciding the outputs based on the inputs by $\tilde{p}_A(a|x,\lambda)\tilde{p}_B(b|y,\lambda)$, and the Bell value with a LHVM is given by
\begin{equation}\label{Eq:Bellvalue}
  J^{\mathrm{LHVM}} = \frac{1}{p(x,y)}\sum_\lambda\sum_{a,b,x,y} \beta_{a,b}^{x,y}\tilde{p}_A(a|x,\lambda)\tilde{p}_B(b|y,\lambda)p(x,y|\lambda)q(\lambda).
\end{equation}
What we are interested is to maximize $J^{\mathrm{LHVM}}(P, Q)$ with LHVMs. From another point of view, we want to establish the Bell's inequality when imperfectly random inputs are considered. Any breach of these bounds (using quantum settings) would rule out LHVMs and in favor of quantum mechanics. Suppose the quantum bound to Eq.~\eqref{Eq:BellFree} is denoted by $J_Q$, then we are especially interested to see what is the condition of $P$ and $Q$ so that $J^{\mathrm{LHVM}}(P, Q) < J_Q$. In experiment, such conditions are the necessary condition for a valid Bell test. For a specific observed violation $J_{\mathrm{obs}}$ and input randomness characteristics $P$ and $Q$, it witnesses non-local feature only if the Bell value satisfies $J^{\mathrm{LHVM}}(P, Q) < J_{\mathrm{obs}}$.

In varieties of previous works \cite{Hall10,Barrett11,Hall11,Koh12,Pope13,Thinh13, Yuan15}, such requirement for the CHSH inequality is analyzed. In this work, we focus on another inequality --- the CH inequality and consider in general scenarios. For instance, many previous work \cite{Hall10,Hall11,Koh12,Pope13, Yuan15} assumes the underlying probability distribution $\tilde{p}_{AB}(a,b|x,y)$ to satisfy the no-signaling (NS) \cite{prbox} condition. However, in real experiment, the probability distribution $\tilde{p}_{AB}(a,b|x,y)$ may behave signaling due to statistical fluctuation, devices imperfection, or other possible interventions by the adversary Eve. We thus also consider the general case where $\tilde{p}_{AB}(a,b|x,y)$ can be signaling. In addition, we consider the case that the random inputs of Alice and Bob are factorizable. In this case, the input randomness can be written as
\begin{equation}\label{Eq:Uncorrelated}
  p(x,y|\lambda) = p_A(x|\lambda)p_B(y|\lambda).
\end{equation}
This factorizable assumption is reasonable in some practical scenarios, where the experiment devices that determine the input settings are from independent manufactures or the randomness generation events are also spacelikely separated. For example, if the inputs are determined by cosmic photons that are causally disconnected from each other \cite{Gallicchio14}, the input randomness can be factorizable.

A more general measure for the input randomness is based on the Santha-Vazirani sources, which is popularly applied in randomness expansion protocols \cite{colbeck2012free}. We note that, the input randomness measure in Eq.~\eqref{eq:randomness} is a special case of the Santha-Vazirani source, where we assume the input settings from different runs are independent and identically distributed (i.i.d.).

\section{CH inequality}\label{Sec:CH}
In this section, we will investigate the randomness requirement of the CH inequality under different conditions, including whether $\tilde{p}_{AB}(a,b|x,y)$ is signaling or NS, and whether the factorizable condition is satisfied or not.
\subsection{CH inequality with LHVMs}
The CH inequality is defined in the bipartite scenario, where the input settings $x$ and $y$ and the outputs $a$ and $b$ are all bits. Based on the probability distribution that obtains a specific measurement outcome, for instance $00$, the CH inequality is defined according to
\begin{equation}\label{eq:CH}
\begin{aligned}
  J_{\mathrm{CH}}  &= \tilde{p}_{AB}(0,0) + \tilde{p}_{AB}(0,1) +\tilde{p}_{AB}(1,0) \\
  &-\tilde{p}_{AB}(1,1) -\tilde{p}_{A}(0) -\tilde{p}_{B}(0)\leq 0,
\end{aligned}
\end{equation}
where we omit the outputs $a$ and $b$ and define $\tilde{p}_{A}(x)$ ($\tilde{p}_B(y)$) to be the probability of detecting $0$ condition on input setting $x$ ($y$) by Alice (Bob), and $\tilde{p}_{AB}(x,y)$ the probability of coincidence detection $00$ for both sides with input settings $x$ and $y$ for Alice and Bob, respectively.
To satisfy the general definition of Bell's inequality as shown in Eq.~\eqref{eq:Bell}, the single party probabilities $\tilde{p}_{A}(0)$ and  $\tilde{p}_{B}(0)$ need to be properly defined by coincidence detection probabilities. For instance, we can either define $p_{A}(0)$ by the detection probabilities with input $(x = 0, y = 0)$, or $(x = 0, y = 1)$, or a convex mixture. This arbitrary definition vanishes when the NS condition is satisfied.

In experiment realization, one has to run the CH test multiple times, for instance, $N$, to determine the probabilities in Eq.~\eqref{eq:CH}. Denote the coincidence counts by $C_{AB}$  and single counts by $S_{A(B)}$, we can then write
\begin{equation}\label{eq:CH2}
\begin{aligned}
  J_{\mathrm{CH}}  &= \frac{C_{AB}(0,0)}{N_{AB}(0,0)} + \frac{C_{AB}(0,1)}{N_{AB}(0,1)} +\frac{C_{AB}(1,0)}{N_{AB}(1,0)}\\
   &-\frac{C_{AB}(1,1)}{N_{AB}(1,1)} -\frac{S_{A}(0)}{N_{A}(0)} -\frac{S_{B}(0)}{N_{B}(0)}.
\end{aligned}
\end{equation}
Here, $N_{AB}(x,y)$ denotes the total number of trials with input setting $x$ and $y$, and $N_{A(B)}$ the number of trials with input setting $x$ ($y$) of Alice (Bob).

When the input settings are chosen truly randomly,  the CH Bell value $J^{\mathrm{LHVM}}_{\mathrm{CH}}$ with LHVM is always non-positive. While quantum theory could maximally violate it to be $J_Q= (\sqrt{2} - 1)/2\approx0.207$.
If the measurement settings $x$, $y$ are additionally determined by some hidden variable $\lambda$ by probability distribution $p(x, y|\lambda)$, we show in the following that the CH inequality could be violated even with LHVMs.

With a general LHVM strategy defined in Eq.~\eqref{Eq:Bellvalue}, each term in the CH value in Eq.~\eqref{eq:CH2} can be described by
\begin{equation}\label{}
\begin{aligned}
C_{AB}(x,y)&=N\sum_\lambda \tilde{p}_{A}(x,\lambda)\tilde{p}_{B}(y,\lambda)p(x,y|\lambda)q(\lambda)\\
N_{AB}(x,y)&=N\sum_\lambda p(x,y|\lambda)q(\lambda)\\
S_{A(B)}(0)&=N\sum_\lambda \tilde{p}_{A(B)}(0,\lambda)\left(p(0,0|\lambda)+p(0,1|\lambda)\right)q(\lambda)\\
N_{A(B)}(0)&=N\sum_\lambda \left(p(0,0|\lambda)+p(0,1|\lambda)\right)q(\lambda)\\
\end{aligned}
\end{equation}
Here, we adopt a specific realization of the single counts by taking an average of the observed value. For instance, the single detection probability $p_{A}(0)$ is defined to be a mean of the single detection probabilities with input $(x = 0, y = 0)$ and $(x = 0, y = 1)$.

Besides, in order to convince Alice and Bob that the input settings $x$ and $y$ are chosen freely, Eve has to impose that the averaged probability distributions of the input settings are uniformly random. Then, we can assume $p(x,y)$ to be $1/4$,
\begin{equation}\label{eq:CHReq}
\begin{aligned}
  N_{AB}(x,y) = N\sum_\lambda p(x,y|\lambda)q(\lambda) =  N/4, \forall x, y.
\end{aligned}
\end{equation}
In real experiments, the input probability can be arbitrary, where our result can still apply with certain modifications on normalization. With the normalization condition Eq.~\eqref{eq:CHReq}, the CH value with LHVMs strategies is given by
\begin{equation}\label{Eq:CHLHVM}
  J^{\mathrm{LHVM}}_{\mathrm{CH}} = 4\sum_{\lambda} q(\lambda) J_\lambda
\end{equation}
with $J_\lambda$  defined by
\begin{equation}\label{}
\begin{aligned}\label{}
  J_\lambda &= \tilde{p}_{A}(0,\lambda)\tilde{p}_{B}(0,\lambda)p(0,0|\lambda) + \tilde{p}_{A}(0,\lambda)\tilde{p}_{B}(1,\lambda)p(0,1|\lambda) \\ &+\tilde{p}_{A}(1,\lambda)\tilde{p}_{B}(0,\lambda)p(1,0|\lambda)
-\tilde{p}_{A}(1,\lambda)\tilde{p}_{B}(1,\lambda)p(1,1|\lambda) \\ &-\tilde{p}_{A}(0,\lambda)(p(0,0|\lambda)+p(0,1|\lambda))/2 \\ &-\tilde{p}_{B}(0,\lambda)(p(0,0|\lambda)+p(1,0|\lambda))/2.
\end{aligned}
\end{equation}
With the randomness parameter defined in Eq.~\eqref{eq:randomness}, our target is to maximize $J^{\mathrm{LHVM}}_{\mathrm{CH}}$ defined in Eq.~\eqref{Eq:CHLHVM} with given randomness input $P$ and $Q$ under constraints in Eq.~\eqref{eq:CHReq}.

\subsection{General strategy (attack)}
In this part, we consider a general strategy (attack) where no additional assumption is imposed. It is worth mentioning that with the following method, we can essentially convert the optimization problem over all LHVMs into a well defined mathematical problem. In the CH example, we show an explicit solution to this mathematical problem. A general solution to this type mathematical problem will provide a solution for the problem with a general Bell's inequality.

Note that the optimization of Eq.~\eqref{Eq:CHLHVM} requires to optimize over the strategy of Alice and Bob, $\tilde{p}_A(x,\lambda)$ and $\tilde{p}_B(y,\lambda)$, and also the strategy of deciding the inputs, $p(x,y|\lambda)$, which also satisfies the constraints defined in Eq.~\eqref{eq:CHReq}. Here, we first analyze how to optimize the strategy of Alice and Bob.

Because all probabilistic LHVM strategies can be realized with a convex combination of deterministic strategies, it is sufficient to just consider deterministic strategies, i.e., $\tilde{p}_A(x),\tilde{p}_B(y)\in\{0,1\}$ for the optimization. Conditioned on different values of $\tilde{p}_A(x)$ and $\tilde{p}_B(y)$, 16 possible values of $J_\lambda$ are listed in Table~\ref{table:CHattack1}, where we omit the $\lambda$ for simple notation hereafter.
\begin{table*}[hbt]\label{table:CHattack1}
\centering
\caption{The value of $J_\lambda$ with deterministic strategy.}
\begin{tabular}{cccccc}
  \hline
  &&\multicolumn{4}{c}{$(\tilde{p}_B(0),\tilde{p}_B(1))$}\\
   &&$(0,0)$&$(0,1)$&$(1,0)$&$(1,1)$\\
   \hline
     \multirow{4}{*}{$(\tilde{p}_A(0),\tilde{p}_A(1))$}&$(0,0)$&$0$&$0$&$-(p(0,0)+p(1,0))/2$&$-(p(0,0)+p(1,0))/2$\\
    &$(0,1)$&$0$&$-p(1,1)$&$(p(1,0)-p(0,0))/2$&$(p(1,0)-p(0,0))/2-p(1,1)$\\
    &$(1,0)$&$-(p(0,0)+p(0,1))/2$&$(p(0,1)-p(0,0))/2$&$-(p(0,1)+p(1,0))/2$&$(p(0,1)-p(1,0))/2$\\
    &$(1,1)$&$-(p(0,0)+p(0,1))/2$&$(p(0,1)-p(0,0))/2-p(1,1)$&$(p(1,0)-p(0,1))/2$&$(p(1,0)+p(0,1))/2-p(1,1)$\\
  \hline
\end{tabular}
\label{table:CHattack1}
\end{table*}
Note that, for given $p(x,y|\lambda)$, we should choose the optimal strategy of $\tilde{p}_A(x)$ and $\tilde{p}_B(y)$ that maximize $J_\lambda$. Thus we here only consider the possible optimal strategies as listed in Table~\ref{table:CHattack2}. We refer to Appendix~\ref{App:proof1} for rigorous proof of why we only consider the possible optimal strategies.

\begin{table}[hbt]\label{table:CHattack2}
\centering
\caption{Possible strategies for letting $J_\lambda$ be positive. }
\begin{tabular}{cc}
  \hline
  $(\tilde{p}_A(0),\tilde{p}_A(1),\tilde{p}_B(0),\tilde{p}_B(1))$&$J_\lambda$\\
  \hline
(0,1,1,0)&$(p(1,0)-p(0,0))/2$\\
(0,1,1,1)&$(p(1,0)-p(0,0))/2-p(1,1)$\\
(1,0,0,1)&$(p(0,1)-p(0,0))/2$\\
(1,0,1,1)&$(p(0,1)-p(1,0))/2$\\
(1,1,0,1)&$(p(0,1)-p(0,0))/2-p(1,1)$\\
(1,1,1,0)&$(p(1,0)-p(0,1))/2$\\
(1,1,1,1)&$(p(1,0)+p(0,1))/2-p(1,1)$\\
  \hline

\end{tabular}
\label{table:CHattack2}
\end{table}

As the strategies of $(\tilde{p}_A(0), \tilde{p}_A(1),\tilde{p}_B(0),\tilde{p}_B(1)) = (0,1,1,0)$ and $(\tilde{p}_A(0), \tilde{p}_A(1),\tilde{p}_B(0),\tilde{p}_B(1)) = (1,0,0,1)$ are always better than the strategies of $(\tilde{p}_A(0), \tilde{p}_A(1),\tilde{p}_B(0),\tilde{p}_B(1)) = (0,1,1,1)$ and $(\tilde{p}_A(0), \tilde{p}_A(1),\tilde{p}_B(0),\tilde{p}_B(1)) = (1,1,0,1)$, respectively, we can always replace the later strategies with the former ones without affecting $p(x,y)$ but achieving a larger $J_\lambda$.
For simple notation, we denote $p(i,j)$ by $p_{2*i + j}$ hereafter, thus the possible deterministic strategies for $J_\lambda$ are in the following set
\begin{equation}\label{eq:chq}
\begin{aligned}
  \left\{\frac{p_2-p_0}{2}, \frac{p_1-p_0}{2}, \frac{p_1-p_2}{2}, \frac{p_2-p_1}{2}, \frac{p_2+p_1}{2}-p_3\right\}.
\end{aligned}
\end{equation}

As there are only five possible strategies of Alice and Bob, we can also consider that there are only five strategies of choosing the input settings. The intuition is that, for the input settings that using the same strategies of Alice and Bob, for instance, $J_\lambda = (p_2 - p_0)/2$, we can always take an average of the different strategies of $p(x,y|\lambda)$ without decreasing $J_\lambda$. We refer to Appendix~\ref{App:proof1} for a rigorous proof.
Therefore, we label $\lambda_j$ to be the $j$th strategy of choosing the input settings and $J^{\mathrm{LHVM}}_{\mathrm{CH}}$ can be rewritten in the following way,
\begin{equation}\label{eq:Proof}
\begin{aligned}
  &J^{\mathrm{LHVM}}_{\mathrm{CH}}/4 \\
  &= q(\lambda_1)(p_2(\lambda_1)-p_0(\lambda_1))/2 + q(\lambda_2)(p_1(\lambda_2)-p_0(\lambda_2))/2\\
  &+q(\lambda_3)(p_1(\lambda_3)-p_2(\lambda_3))/2+q(\lambda_4)(p_2(\lambda_4)-p_1(\lambda_4))/2\\
  &+q(\lambda_5)[(p_2(\lambda_5)+p_1(\lambda_5))/2-p_3(\lambda_5)].
\end{aligned}
\end{equation}
The constraints of $q(\lambda)$ and $p(\lambda)$ are given by
\begin{equation}\label{eq:constraints}
\begin{aligned}
  &\sum_j q(\lambda_j) p_i(\lambda_j) = 1/4, \forall i \\
  &\sum_i p_i(\lambda_j) = 1, \forall j, \\
  &\sum_j q(\lambda_j) = 1,\\
   &Q\le p_i(\lambda_j)\le P, \forall i,j.\\
\end{aligned}
\end{equation}

Furthermore, we can denote the coefficient of $q(\lambda_j)p_i(\lambda_j)$ by $\beta_{ij}$ as shown in Table~\ref{Table:coe}. Then $J^{\mathrm{LHVM}}_{\mathrm{CH}}$ can be expressed by
\begin{equation}\label{Eq:mathJlambda}
  J^{\mathrm{LHVM}}_{\mathrm{CH}} = 4\sum_{ij} \beta_{ij} q(\lambda_j)p_i(\lambda_j),
\end{equation}

\begin{table}[hbt]
\centering
\caption{The coefficient $\beta_{ij}$ of $q(\lambda_j)p_i(\lambda_j)$ in the expression of $J^{\mathrm{LHVM}}_{\mathrm{CH}}$ of the CH inequality.}
\begin{tabular}{cccccc}
  \hline
   &$\lambda_1$&$\lambda_2$&$\lambda_3$&$\lambda_4$&$\lambda_5$\\
   \hline
   $p_0$&$-\frac{1}{2}$&$\frac{1}{2}$&$0$&$0$&$0$\\
   $p_1$&$0$&$\frac{1}{2}$&$\frac{1}{2}$&$-\frac{1}{2}$&$\frac{1}{2}$\\
   $p_2$&$\frac{1}{2}$&$0$&$-\frac{1}{2}$&$\frac{1}{2}$&$\frac{1}{2}$\\
   $p_3$&$0$&$0$&$0$&$0$&$-1$\\
  \hline
\end{tabular}\label{Table:coe}
\end{table}

\begin{figure}[!ht]
    \centering
 \resizebox{8cm}{!}{\includegraphics[scale=1]{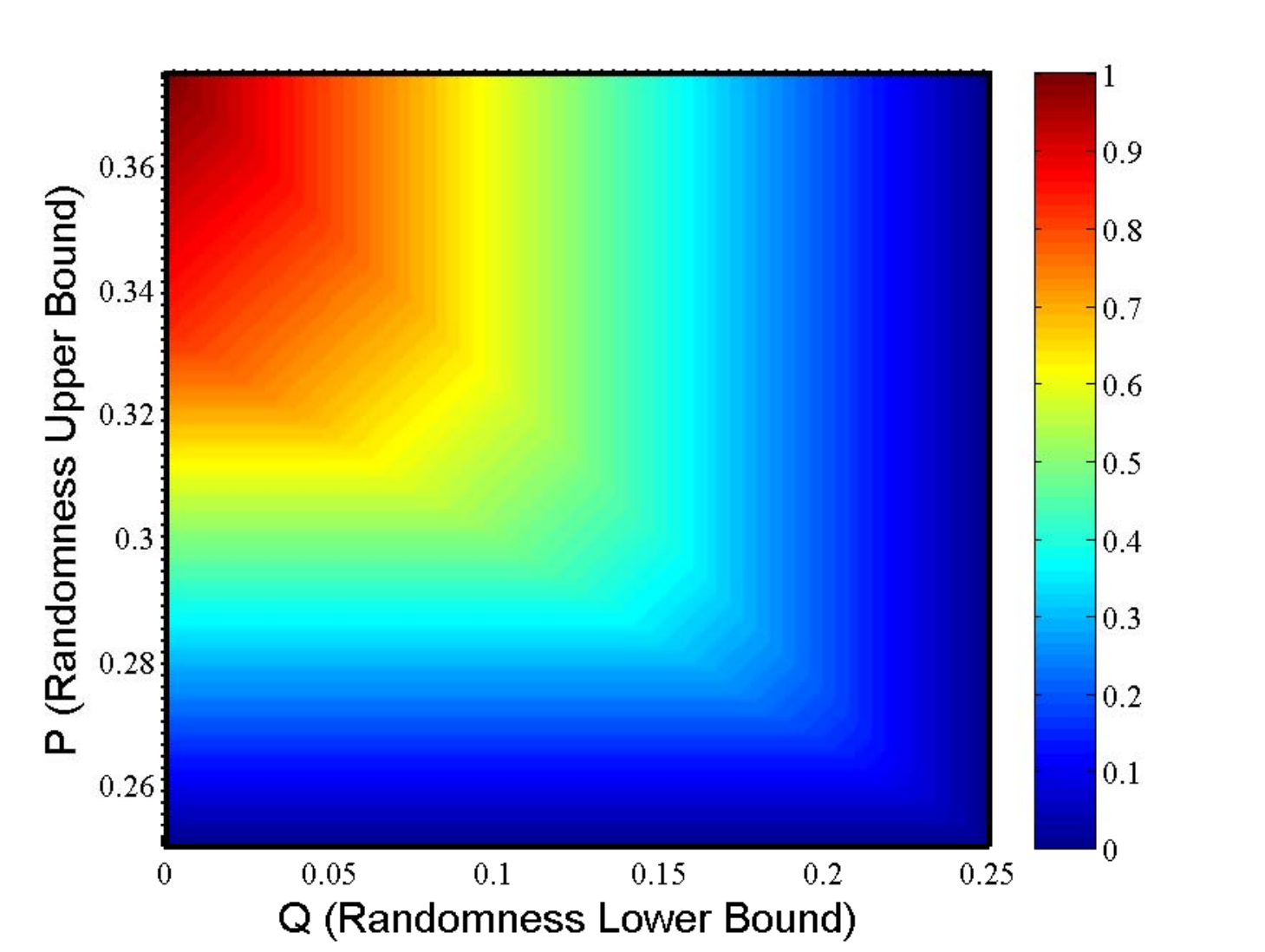}}
\caption{(Color online) The CH value $J^{\mathrm{LHVM}}_{\mathrm{CH}}$ as a function of $P$ and $Q$, according to Eq.~\eqref{eq:JCH}. }\label{Fig:plot0}
\end{figure}

The solution to this optimization problem is shown in Appendix~\ref{app:CH1}. Based on the value of $P$ and $Q$, we give the optimal CH value $J^{\mathrm{LHVM}}_{\mathrm{CH}}$ with LHVMs by
\begin{equation}\label{eq:JCH}
  J^{\mathrm{LHVM}}_{\mathrm{CH}}(P,Q) =
  \left\{
  \begin{array}{cc}
    \frac{5}{2}(4P-1) & 3P+Q\le 1,\\
    1-4Q & 2P+Q\ge \frac{3}{4},\\
     4P-2Q-\frac{1}{2} & \mathrm{else},
  \end{array}
  \right.
\end{equation}
and plot it in Fig.~\ref{Fig:plot0}. Note that when $P$ is greater than ${3}/{8}$, the value of $J^{\mathrm{LHVM}}_{\mathrm{CH}}$ is independent of $P$. Hence, we only plot the situation where $P$ is less than 3/8.

In addition, we can also investigate the optimal CH value $J^{\mathrm{LHVM}}_{\mathrm{CH}}$ with input randomness quantified as in Eq.~\eqref{Eq:deltasource}. It is easy to check that $2P + Q\geq 3/4$, and the optimal CH value $J^{\mathrm{LHVM}}_{\mathrm{CH}}$ is thus
\begin{equation}\label{}
  J^{\mathrm{LHVM}}_{\mathrm{CH}}(\delta) = 4\delta.
\end{equation}

\subsection{factorizable condition}
Here, we consider the optimal LHVMs strategy in the case where the probability of the input settings are  factorizable, as defined in Eq.~\eqref{Eq:Uncorrelated}.

\begin{figure}[!ht]
    \centering
 \resizebox{8cm}{!}{\includegraphics[scale=1]{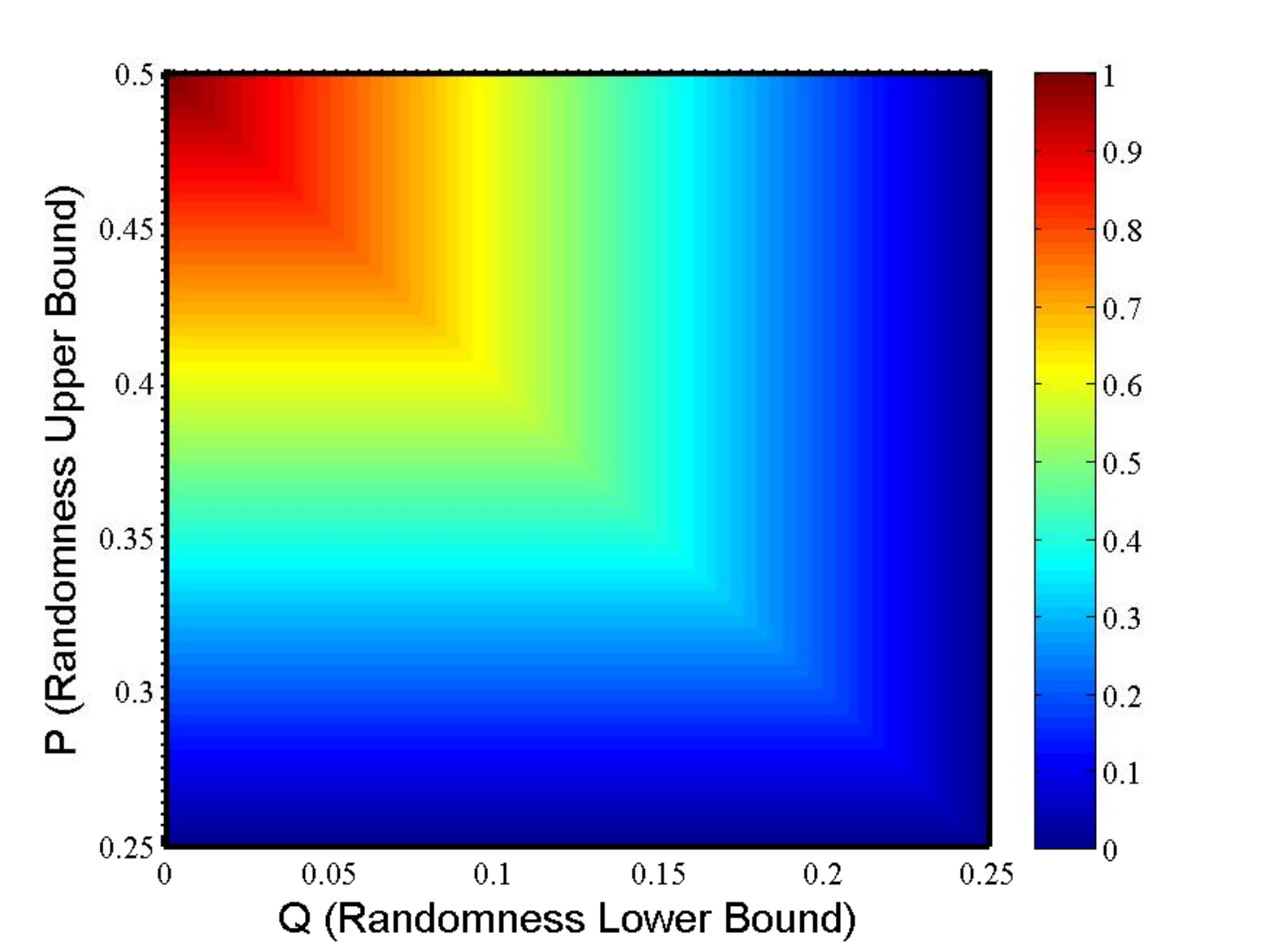}}
\caption{(Color online) The CH value $J^{\mathrm{LHVM, Fac}}_{\mathrm{CH}}$ as a function of $P$ and $Q$ with the factorizable condition Eq.~\eqref{Eq:Uncorrelated}. }\label{Fig:plot1}
\end{figure}

Following a similar derivation, we show in the Appendix~\ref{app:fac} that the optimal CH value $J^{\mathrm{LHVM, Fac}}_{\mathrm{CH}}$ with LHVMs under factorizable condition is
\begin{equation}\label{}
  J^{\mathrm{LHVM, Fac}}_{\mathrm{CH}}(P,Q) =
  \left\{
  \begin{array}{cc}
    (4P-1) & P+Q\le \frac{1}{2},\\
    1-4Q & P+Q> \frac{1}{2}.\\
  \end{array}
  \right.
\end{equation}
We show the optimal value of  $J^{\mathrm{LHVM, Fac}}_{\mathrm{CH}}$  in Fig.~\ref{Fig:plot1}.

When we quantify $P$ and $Q$ by $P = 1/4 + \delta$ and $Q = 1/4 - \delta$, the fomular can be rewritten by
\begin{equation}\label{}
  J^{\mathrm{LHVM, Fac}}_{\mathrm{CH}}(\delta) = 4\delta.
\end{equation}

It is interesting to note that the factorizable condition does not affect the optimal CH value $J^{\mathrm{LHVM}}_{\mathrm{CH}}(\delta)$ when the input randomness is quantified as in Eq.~\eqref{Eq:deltasource}. The quantum bound $J_Q$ is given by $(\sqrt{2}-1)/2$, thus we can see that $\delta$ should  at least be less than $0.051$ for all CH experiment realizations.

\subsection{NS condition}\label{Sec:CHSH}
In addition, we consider the scenario where the probability distribution $\tilde{p}_{AB}(a,b|x,y)$ defined in Eq.~\eqref{Eq:pabxy} satisfies the NS condition, which adds a constraint on $\tilde{p}_{AB}(a,b|x,y)$. That is, the probability of output $a$ ($b$) only relies on the input $x$ ($y$) independently of the input from the other party. To be more specific, NS requires $\tilde{p}_{AB}(a,b|x,y)$ to satisfy
 \begin{equation}\label{eq:nosignaling}
 \begin{aligned}
&\sum_b \tilde{p}_{AB}(a,b|x,y)=\sum_b \tilde{p}(a,b|x,y')\equiv \tilde{p}_A(a|x), ~ \forall a,x,y,y'\\
&\sum_a \tilde{p}_{AB}(a,b|x,y)=\sum_a \tilde{p}(a,b|x',y)\equiv \tilde{p}_B(b|y). ~\forall b,x,x',y
\end{aligned}
\end{equation}
We can follow the above derivation by imposing an additional NS constraint, which makes the problem even more complex.

Instead, we note that the CHSH inequality and the CH inequality are equivalent under NS, that is,
\begin{equation}\label{Eq:CHCHSH}
  J_{\mathrm{CH}}^{\mathrm{NS}} = \frac{1}{4}(J_{\mathrm{CHSH}} - 2),
\end{equation}
which we refer to Appendix~\ref{app:equiva} for a rigorous proof. As the CHSH inequality is defined with strong symmetry, we  solve the optimization problem with the CHSH inequality. We should note that we essentially take the NS condition into account when deriving the equivalence between the CH and CHSH inequality.

Based on the general definition of Bell's inequalities in Eq.~\eqref{eq:Bell}, the coefficients of the CHSH inequality is defined by
\begin{equation}\label{}
    \beta_{a,b,x,y}^{\mathrm{CHSH}} = (-1)^{xy + a + b},
\end{equation}
that is,
\begin{equation}\label{}
  J_{\mathrm{CHSH}} = \sum_{x,y,a,b\in\{0,1\}}(-1)^{xy + a + b}\tilde{p}(a, b|x, y).
\end{equation}

When considering LHVMs strategies with imperfect input randomness, the CHSH value can be written by
\begin{equation}\label{}
    J_{\mathrm{CHSH}}^{\mathrm{LHVM}} = 4\sum_\lambda q_\lambda J_\lambda,
\end{equation}
where
\begin{equation}\label{}
  J_\lambda = \sum_{x,y,a,b}(-1)^{xy + a + b}\tilde{p}_A(a|x,\lambda)\tilde{p}_B(b|y,\lambda)p(x,y|\lambda).
\end{equation}

Following a similar method described above, we first consider deterministic strategies, i.e., $\tilde{p}_A(a|x),\tilde{p}_B(b|y)\in\{0,1\}$ for the reason that any probabilistic LHVM could be realized with convex combination of deterministic ones.
Denote $p(i,j)$ as $p_{2*i + j}$, it is easy to show that the possible optimal deterministic strategies for $J_\lambda$ are
\begin{equation}
\begin{aligned}
   \{p_0+p_1+p_2-p_3&, p_0+p_1+p_3-p_2, \\
   p_0+p_2+p_3-p_1&, p_1+p_2+p_3-p_0\},
\end{aligned}
\end{equation}
and the constraints can also be described by Eq.~\eqref{eq:constraints}. Following a similar argument, we only need to consider four different types strategies of choosing the input settings. Thus $J^{\mathrm{LHVM}}_{\mathrm{CHSH}}$ can be given by
\begin{equation}\label{chsh}
\begin{aligned}
  &J^{\mathrm{LHVM}}_{\mathrm{CHSH}}/4 \\
  &= q(\lambda_1)(p_0(\lambda_1)+p_1(\lambda_1)+p_2(\lambda_1)-p_3(\lambda_1))\\
  &+q(\lambda_2)(p_0(\lambda_2)+p_1(\lambda_2)+p_3(\lambda_2)-p_2(\lambda_2))\\
  &+q(\lambda_3)(p_0(\lambda_3)+p_2(\lambda_3)+p_3(\lambda_3)-p_1(\lambda_3))\\
  &+q(\lambda_4)(p_1(\lambda_4)+p_2(\lambda_4)+p_3(\lambda_4)-p_0(\lambda_4)).
\end{aligned}
\end{equation}
With a symbolic notation in Eq.~\eqref{Eq:mathJlambda}, we can also present the coefficient $\beta_{ij}$ in a matrix, as shown in Table~\ref{Table:coeCHSH}.

\begin{table}[hbt]
\centering
\caption{The coefficient of $p_i(\lambda_j)$ in the expression of $J^{\mathrm{LHVM}}_{\mathrm{CHSH}}$ of the CHSH inequality.}
\begin{tabular}{ccccc}
  \hline
   &$\lambda_1$&$\lambda_2$&$\lambda_3$&$\lambda_4$\\
   \hline
   $p_0$&$1$&$1$&$1$&$-1$\\
   $p_1$&$1$&$1$&$-1$&$1$\\
   $p_2$&$1$&$-1$&$1$&$1$\\
   $p_3$&$-1$&$1$&$1$&$1$\\
  \hline
\end{tabular}\label{Table:coeCHSH}
\end{table}

\begin{figure}[!ht]
    \centering
 \resizebox{8cm}{!}{\includegraphics[scale=1]{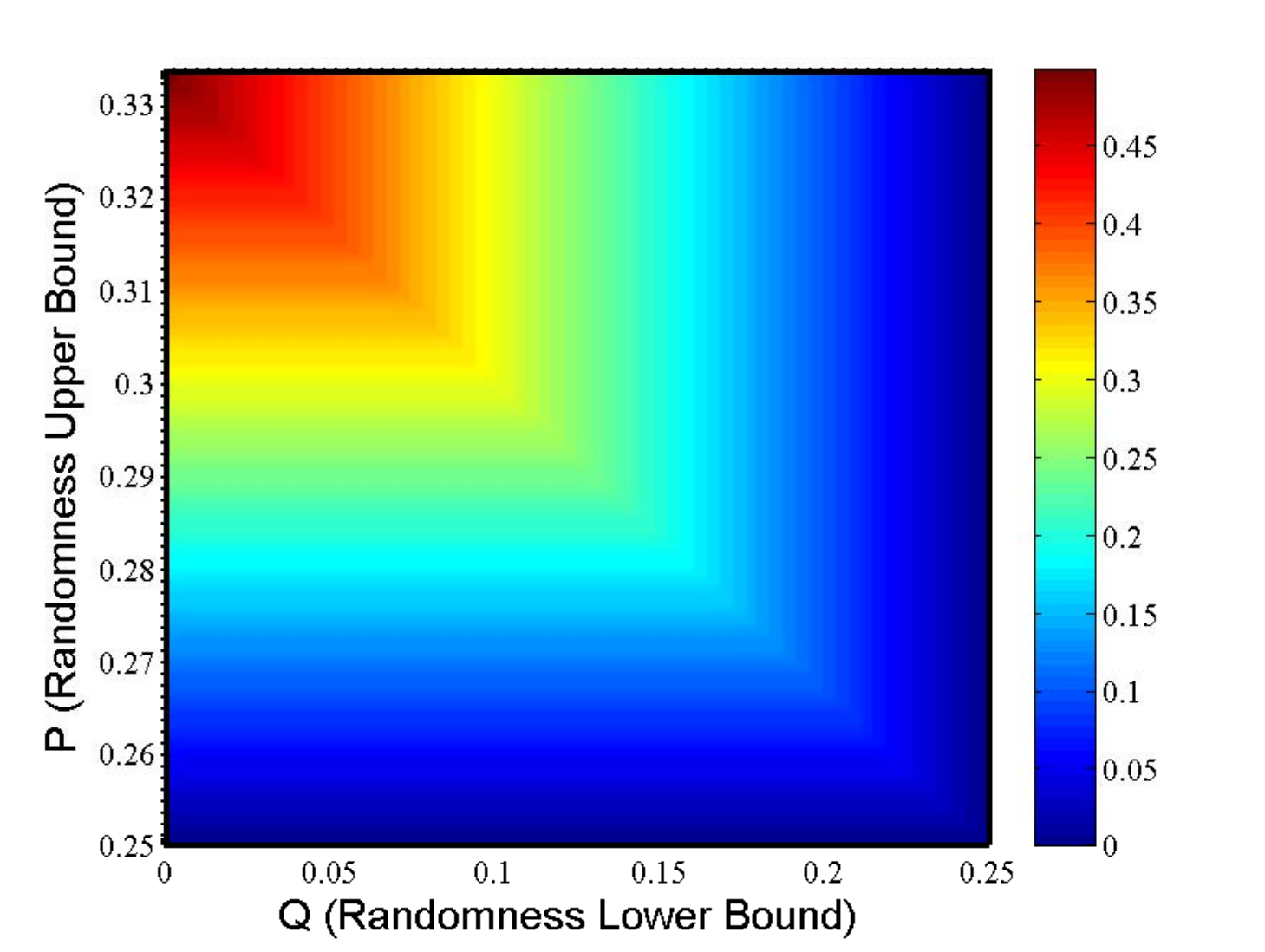}}
\caption{(Color online) The CH value $J^{\mathrm{LHVM, \mathrm{NS}}}_{\mathrm{CH}}$ as a function of $P$ and $Q$ under NS condition Eq.~\eqref{eq:nosignaling}. }\label{Fig:plot2}
\end{figure}

We solve the optimization problem in Appendix~\ref{app:CHSH}. Based on the value of $P$ and $Q$, we give the optimal CHSH value $J^{\mathrm{LHVM}}_{\mathrm{CHSH}}$ with LHVMs by
\begin{equation}\label{}
  J^{\mathrm{LHVM}}_{\mathrm{CHSH}}(P,Q) =
  \left\{
  \begin{array}{cc}
    24P-4 & 3P+Q\le 1,\\
    4-8Q & 3P+Q\ge 1.
  \end{array}
  \right.
\end{equation}
Then the optimal CH value $J^{\mathrm{LHVM, NS}}_{\mathrm{CH}}$ with LHVMs under NS is
\begin{equation}\label{}
  J^{\mathrm{LHVM, \mathrm{NS}}}_{\mathrm{CH}}(P,Q) =
  \left\{
  \begin{array}{cc}
    6P-3/2 & 3P+Q\le 1,\\
    1/2-2Q & 3P+Q\ge 1.
  \end{array}
  \right.
\end{equation}
We show the optimal value of  $J^{\mathrm{LHVM, \mathrm{NS}}}_{\mathrm{CH}}$  in Fig.~\ref{Fig:plot2}.

If we quantify the input randomness by its deviation from uniform distribution as defined in Eq.~\eqref{Eq:deltasource}, the optimal CH value $J^{\mathrm{LHVM,NS}}_{\mathrm{CH}}(\delta)$  is given by
\begin{equation}\label{}
  J^{\mathrm{LHVM}}_{\mathrm{CH,NS}}(\delta) = 2\delta.
\end{equation}
The quantum bound for the CH inequality $J_Q$ is $(\sqrt{2}-1)/2$, thus  $\delta$ should be less than $(\sqrt{2} - 1)/4\approx 0.104$ for all experiment realizations.

\subsection{NS condition and factorizable}
\begin{figure}[!ht]
    \centering
 \resizebox{8cm}{!}{\includegraphics[scale=1]{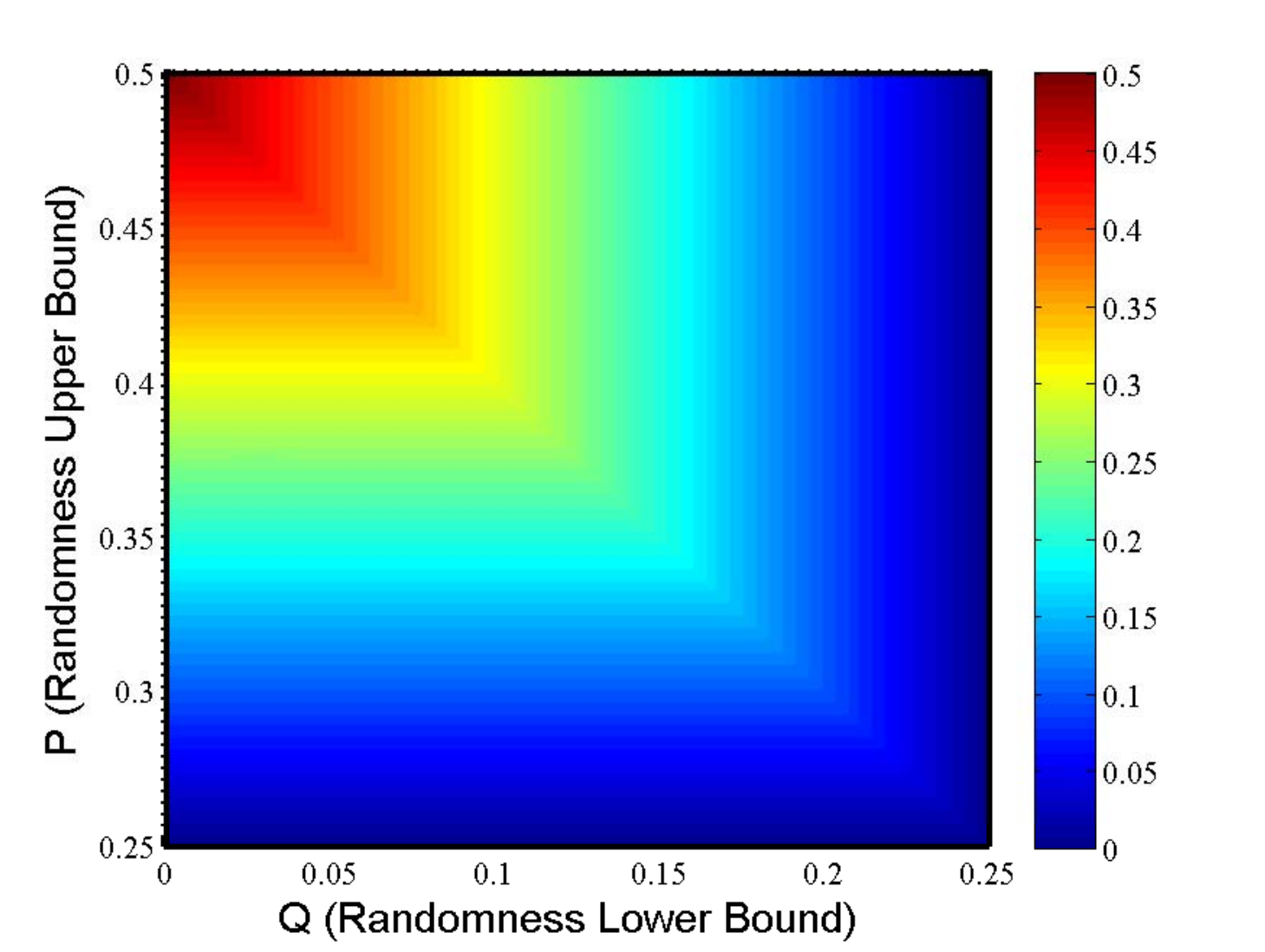}}
\caption{(Color online) The CH value $J^{\mathrm{LHVM, NS, Fac}}_{\mathrm{CH}}$ as a function of $P$ and $Q$ under factorizable Eq.~\eqref{Eq:Uncorrelated} and NS Eq.~\eqref{eq:nosignaling} conditions.} \label{Fig:plot3}
\end{figure}

At last, we consider the probability distribution $\tilde{p}_{AB}(a,b|x,y)$ to be NS and the input randomness $p(x,y|\lambda)$ is factorizable. The optimization of the CHSH inequality is solved in Appendix~\ref{app:NSFac}, and the result is,
\begin{equation}\label{}
  J^{\mathrm{LHVM,Fac}}_{\mathrm{CHSH}}(P,Q) =
  \left\{
  \begin{array}{cc}
    8P & P+Q \le \frac{1}{2},\\
    4-8Q & P+Q> \frac{1}{2}.\\
  \end{array}
  \right.
\end{equation}
Then the optimal CH value $J^{\mathrm{LHVM, NS, Fac}}_{\mathrm{CH}}$ with LHVMs under NS and factorizable condition is
\begin{equation}\label{}
  J^{\mathrm{LHVM, NS, Fac}}_{\mathrm{CH}}(P,Q) =
  \left\{
  \begin{array}{cc}
    2P-1/2 & P+Q \le \frac{1}{2},\\
    1/2-2Q & P+Q> \frac{1}{2}.\\
  \end{array}
  \right.
\end{equation}
We show the optimal value of  $J^{\mathrm{LHVM, NS, Fac}}_{\mathrm{CH}}$  in Fig.~\ref{Fig:plot3}.

When the input randomness is quantified as in Eq.~\eqref{Eq:deltasource}, where $P = 1/4+\delta$ and $Q = 1/4-\delta$, we have
\begin{equation}\label{}
  J^{\mathrm{LHVM, NS, Fac}}_{\mathrm{CH}}(\delta) = 2\delta.
\end{equation}
Again, it is interesting to note that the factorizable condition does not affect the optimal CH value $J^{\mathrm{LHVM,NS}}_{\mathrm{CH}}(\delta)$ when the input randomness is quantified as in Eq.~\eqref{Eq:deltasource}.

\subsection{Results}
\begin{figure}[!ht]
    \centering
    \resizebox{8cm}{!}{\includegraphics[scale=1]{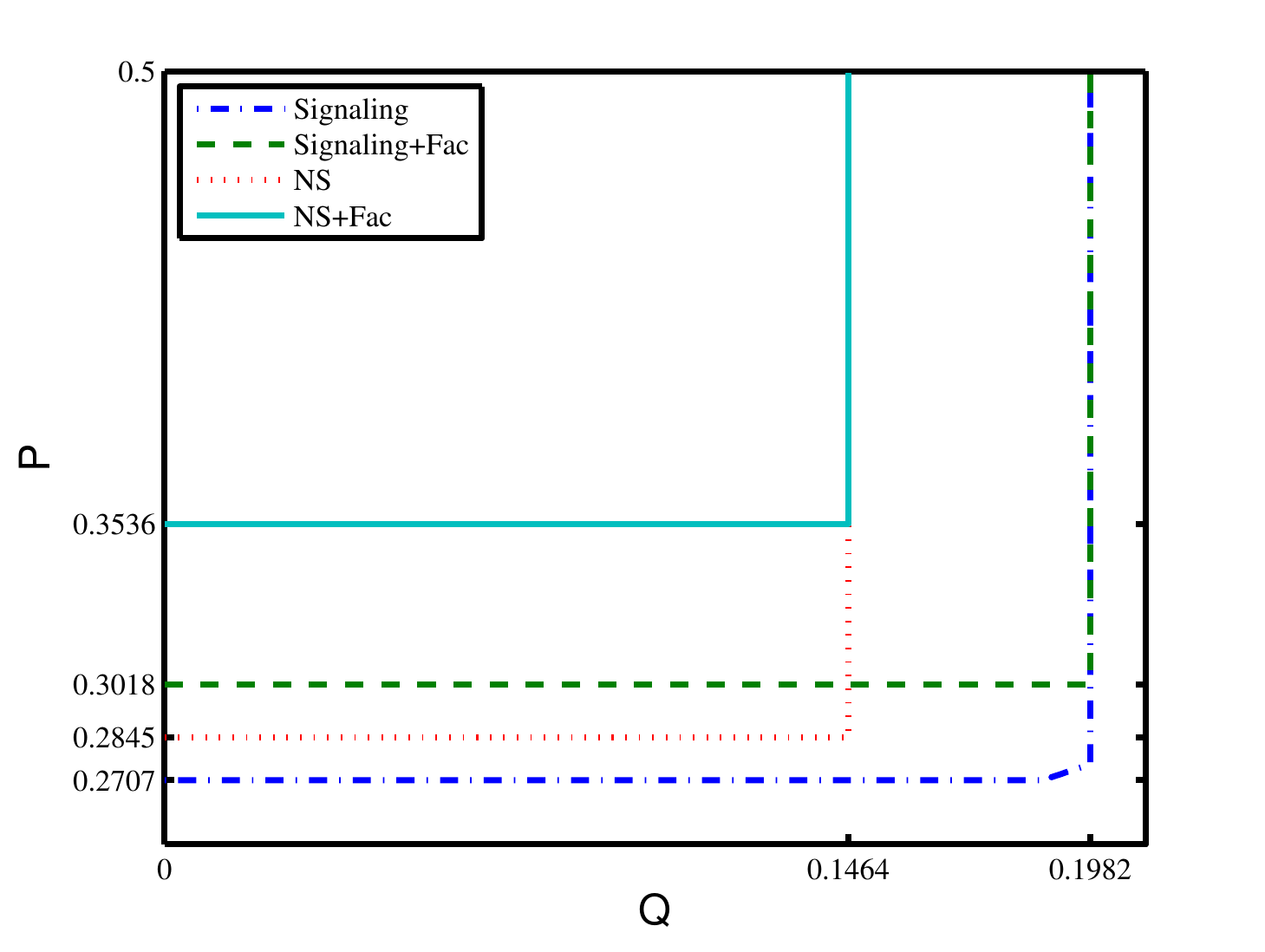}}
    \caption{The critical value of $Q$ and $P$ such that the CH value $J^{\mathrm{LHVM}}_{\mathrm{CH}}(P,Q)$ equals the maximal quantum value $J_Q=(\sqrt{2}-1)/2$.}\label{Fig:PQ}
\end{figure}

Let us compare the results of the CH values $J^{\mathrm{LHVM}}_{\mathrm{CH}}$ under different conditions. For the maximal quantum violation $J_Q=(\sqrt{2}-1)/2$, we calculate the critical values of $Q$ and $P$ such that $J^{\mathrm{LHVM}}_{\mathrm{CH}}(P, Q)=J_Q$ and plot them in Fig.~\ref{Fig:PQ}. When $Q$ is small, the optimal CH value $J^{\mathrm{LHVM}}_{\mathrm{CH}}(P, Q)$ depends only on $P$. In this case, the critical values of $P$ for the signaling, signaling+fac, NS, and NS+fac are 0.207, 0.302, 0.285, 0.356, respectively. Thus, we can see that the factorizable condition puts a stronger requirement for $P$ compared to the NS condition. On the other hand, when $Q$ is large, the optimal CH value $J^{\mathrm{LHVM}}_{\mathrm{CH}}(P, Q)$ depends only on $Q$ instead. In this case, the critical values of $Q$ for the signaling and NS condition are 0.1982 and 0.1464, respectively. It is interesting to note that when both $P$ and $Q$ is large, the optimal CH value $J^{\mathrm{LHVM}}_{\mathrm{CH}}(P, Q)$ is independent on the factorizable condition.

\begin{figure}[!ht]
    \centering
    \resizebox{8cm}{!}{\includegraphics[scale=1]{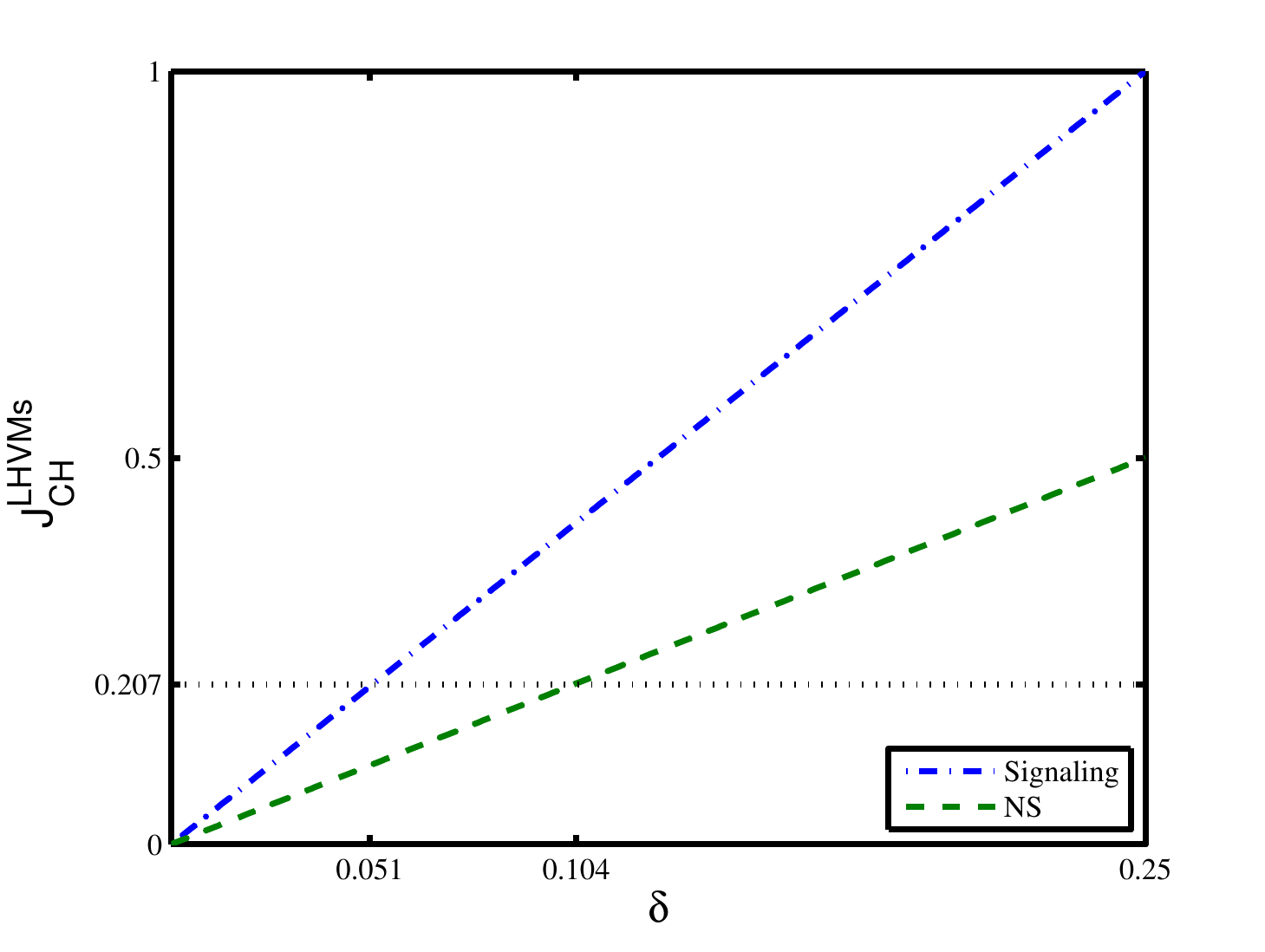}}
    \caption{The CH value $J^{\mathrm{LHVM}}_{\mathrm{CH}}(\delta)$ under different conditions.}\label{Fig:Jdelta}
\end{figure}

Besides, if we make use of the quantification method defined in Eq.~\eqref{Eq:deltasource}, we have already noticed that the optimal CH value $J^{\mathrm{LHVM}}_{\mathrm{CH}}(\delta)$ is independent on the factorizable condition. Here, we compare $J^{\mathrm{LHVM}}_{\mathrm{CH}}(\delta)$ between the signaling and NS condition as shown in Fig.~\ref{Fig:Jdelta}. For the maximal quantum violation $J_Q=(\sqrt{2}-1)/2$, we calculate the critical values of $\delta$ for the signaling and NS condition to be 0.051 and 0.104, respectively.

\section{Conclusion}\label{Sec:conclusion}
In this work, we investigate the randomness requirement in achieving a faithful Bell's test. We explicitly give the optimal LHVM strategy with imperfect randomness to maximize a violation of the CH inequality under different conditions, including whether the strategy is signaling or not and whether the input randomness satisfy the factorizable condition or not. For any observed CH violation, our result thus put an requirement on the input randomness such that the violation is caused by quantum instead of classical process.

When we quantify the input randomness as Eq.~\eqref{Eq:deltasource}, we found that the optimal CH value is independent of the factorizable condition, Eq.~\eqref{Eq:Uncorrelated}. Thus with the quantification method in Eq.~\eqref{Eq:deltasource} one does not need to consider the factorizable condition.
When the NS condition is assumed, the critical value of $\delta$ for the maximal quantum violation is found to be $0.104$. In experiment, such requirement of input randomness is easier to realize compared to the case when only the upper bound $P$ is considered.

For further works, it is interesting to consider joint strategies of LHVMs, where the inputs of different runs are correlated. It is already shown that joint attacks to the CHSH inequality puts a very high requirement of the input randomness no matter the factorizable condition is satisfied or not \cite{Yuan15}. In addition, we can investigate the case where the input randomness is restricted by both lower and upper bounds, $P$ and $Q$. 

Furthermore, it is interesting to see whether there exist Bell's inequalities such that the randomness requirement is very low. To do so, we have to solve the problem of optimizing the Bell value with all LHVM strategies. We expect that our derivation method could provide a general way to solve this problem.

\begin{acknowledgments}
The author acknowledges insightful discussions with Z.~Cao and Z.~Zhang. This work was supported by the National Basic Research Program of China Grants No.~2011CBA00300 and No.~2011CBA00301, and the 1000 Youth Fellowship program in China.
\end{acknowledgments}

\appendix
\section{Proof for finite strategies of choosing input settings}\label{App:proof1}
As we mentioned in the main context, there are two levels of strategies. One is the strategy of choosing the input settings and the other is about the outputs conditioned on inputs of Alice and Bob. As there are finite deterministic strategies of Alice and Bob, here, we prove that the strategies of choosing input settings is finite and can be characterized by all the possible optimal strategies of Alice and Bob.

Essentially, even the strategies of Alice and Bob are finite, the strategies of choosing input settings can always be infinite. Here, what want to prove is that any optimal strategy (including both levels) can be realized with finite strategies of choosing input settings.

Suppose there exist an optimal strategy that gives maximal CH value with LHVMs. For this strategy, we suppose there are finite strategies of choosing the input settings (the proof for infinite case follows similarly). Then, it is easy to check that for a given $\lambda$ and hence $(p_0(\lambda), p_1(\lambda), p_2(\lambda), p_3(\lambda))$  in the optimal strategy, the optimal strategy for the output of Alice and Bob should be from the set  Eq.~\eqref{eq:chq}. This also proves why we only take account of the possibly optimal deterministic strategies of Alice and Bob.

Now, suppose that there exist $m$ strategies of $\lambda$ of choosing input settings for the first strategy of Alice and Bob, $(p_2-p_0)/2$, that is,
\begin{equation}\label{}
  \begin{array}{ccc}
    \lambda_{1}^1 :& q(\lambda_{1}^1),&(p_0(\lambda_{1}^1), p_1(\lambda_{1}^1), p_2(\lambda_{1}^1), p_3(\lambda_{1}^1)), \\
    \lambda_{1}^2 :& q(\lambda_{1}^2),&  (p_0(\lambda_{1}^2), p_1(\lambda_{1}^2), p_2(\lambda_{1}^2), p_3(\lambda_{1}^2)), \\
     & \dots \\
    \lambda_{1}^m :& q(\lambda_{1}^m),&  (p_0(\lambda_{1}^m), p_1(\lambda_{1}^m), p_2(\lambda_{1}^m), p_3(\lambda_{1}^m)).
  \end{array}
\end{equation}
Here the superscript denotes the $m$ strategies of $\lambda$ and the subscript denotes the strategy for Alice and Bob.
It is easy to see that we can always take an average of all the $m$ strategies without decreasing the Bell value and violate the constraints. In this case, we can define one $\lambda_1$ to the denote all the $\lambda_1^1$, $\lambda_1^2$, $\dots\lambda_1^m$. That is,
\begin{equation}\label{}
  \lambda_1: q(\lambda_1) = \sum_{t=1}^mq(\lambda_1^t), (p_0(\lambda_{1}), p_1(\lambda_{1}), p_2(\lambda_{1}), p_3(\lambda_{1})),
\end{equation}
where
\begin{equation}\label{}
  p_i(\lambda_1) = \frac{1}{q(\lambda_1)}\sum_{t=1}^mq(\lambda_1^t)p_i(\lambda_1^t), \forall i \in\{0,1,2,3\}.
\end{equation}
Thus, we show that the $m$ strategies for choosing input settings can be combined into one for any strategy of Alice and Bob. In the following, we prove this argument in more detail.

\begin{proof}
We use label $t$ to denote the $t$th strategy of choosing input settings for a given strategy of Alice and Bob, $j$ to denote the strategies of Alice and Bob, and $i$  to denote the number of inputs meaning the subscript of $(p_0(\lambda), p_1(\lambda), p_2(\lambda), p_3(\lambda))$.

We denote $\lambda_j^t$ to be the $t$th strategy of choosing input settings when the optimal strategy for Alice and Bob is $j$.
The prior probability for $\lambda$ and input settings of each strategy are denoted as $q(\lambda_j^t)$ and $p_i(\lambda_j^t)$, where $j\in\{1,2,3,4,5\}$ and $i\in \{0,1,2,3\}$, respectively. Denote the Bell value for the $j$th strategy to be $J_j$, which is linear function of $\{p_i(\lambda_j^t)\}$. Thus, the total Bell value is given by
\begin{equation}\label{}
  J = \sum_j \sum_t q(\lambda_j^t)J_j(p_i(\lambda_j^t)),
\end{equation}
and the constraints of $q(\lambda_j^t)$ and $p_i(\lambda_j^t)$ are given by,
\begin{equation}\label{infinite}
\begin{aligned}
  &\sum_{j,t} q(\lambda_j^t) p_i(\lambda_j^t) = 1/4, \forall i \\
  &\sum_i p_i(\lambda_j^t) = 1, \forall j,t \\
  &\sum_{j,t} q(\lambda_j^t) = 1.\\
\end{aligned}
\end{equation}

Just as mentioned above, we can add up $t$ by defining $q(\lambda_{j})$ and $p_i(\lambda_{j})$ by
\begin{equation}\label{normalize}
\begin{aligned}
  &q(\lambda_{j})=\sum_{t} q(\lambda_j^t),\forall j\\
  &p_i(\lambda_{j}) = \frac{\sum_t q(\lambda_j^t)p_i(\lambda_j^t)}{q(\lambda_{j})},\forall i,j.\\
\end{aligned}
\end{equation}
Take Eq.~\ref{normalize} into the Eq.~\ref{infinite}, consequently we find the constraints of $q(\lambda_{j})$ and $p_i(\lambda_{j})$ are given by
\begin{equation}\label{}
\begin{aligned}
 &\sum_{j} q(\lambda_{j}) p_i(\lambda_{j}) = 1/4, \forall i \\
  &\sum_i p_i(\lambda_{j}) = 1, \forall j \\
  &\sum_{j} q(\lambda_{j}) = 1.\\
\end{aligned}
\end{equation}

We should also note that the substitution in Eq.~\eqref{normalize} will not affect the Bell value,
\begin{equation}\label{}
\begin{aligned}
  J &= \sum_j q(\lambda_j)J_j(p_i(\lambda_j)),\\
  &=\sum_j  q(\lambda_j)J_j\left(\frac{\sum_t q(\lambda_j^t)p_i(\lambda_j^t)}{q(\lambda_{j})}\right),\\
  & = \sum_j \sum_t q(\lambda_j^t)J_j(p_i(\lambda_j^t)),
\end{aligned}
\end{equation}
where the last equality is because $J_j$ is a  linear function.

\end{proof}

\section{Optimal strategy of the CH test: General condition}\label{app:CH1}
In this section, we present the optimal strategy in order to maximizing $J_{\mathrm{CH}}^{\mathrm{LHVM}}$ defined in Eq.~\eqref{Eq:mathJlambda} under constraints defined in Eq.~\eqref{eq:constraints}.
\subsection{$Q=0$}\label{app:P}
For simplicity, we first consider the randomness requirement $P$ and set $Q$ to be 0. That is, the input randomness is upper bounded by $P$,
\begin{equation}\label{}
  0\leq p(x,y|\lambda) \leq P, \forall x,y,\lambda
\end{equation}

The Bell value $J_{\mathrm{CH}}^{\mathrm{LHVM}}$ with LHVMs is given by
\begin{equation}\label{Eq:appJ}
\begin{aligned}
  &J^{\mathrm{LHVM}}_{\mathrm{CH}}\\
  &= 4\{q(\lambda_1)(p_2(\lambda_1)-p_0(\lambda_1))/2 + q(\lambda_2)(p_1(\lambda_2)-p_0(\lambda_2))/2\\
  &+q(\lambda_3)(p_1(\lambda_3)-p_2(\lambda_3))/2+q(\lambda_4)(p_2(\lambda_4)-p_1(\lambda_4))/2\\
  &+q(\lambda_5)[(p_2(\lambda_5)+p_1(\lambda_5))/2-p_3(\lambda_5)]\}.
\end{aligned}
\end{equation}
Hereafter, we denote  $J_{\mathrm{CH}}^{\mathrm{LHVM}}$ by $J$ for simple notation. Group $J$
by the index of the strategies of Alice and Bob $p_i$, $i\in\{0,1,2,3\}$, instead of $\lambda_i$, then we have
\begin{equation}\label{eq:Proof2}
\begin{aligned}
  J &= 4(J_0 + J_1 + J_2 + J_3),
\end{aligned}
\end{equation}
where
\begin{equation}\label{eq:Proof3}
\begin{aligned}
  J_0 =& \frac{1}{2}[-q(\lambda_1)p_0(\lambda_1) -q(\lambda_2)p_0(\lambda_2)],\\
  J_1 =& \frac{1}{2}[q(\lambda_2)p_1(\lambda_2) +q(\lambda_3)p_1(\lambda_3) \\
  &- q(\lambda_4)p_1(\lambda_4) + q(\lambda_5)p_1(\lambda_5)],\\
  J_2 =& \frac{1}{2}[q(\lambda_1)p_2(\lambda_1) -q(\lambda_3)p_2(\lambda_3) \\
  &+ q(\lambda_4)p_2(\lambda_4) + q(\lambda_5)p_2(\lambda_5)],\\
  J_3 =& -q(\lambda_5)p_3(\lambda_5).\\
\end{aligned}
\end{equation}
And the constraints are given by
\begin{equation}\label{eq:appconstraints}
\begin{aligned}
  &\sum_j q(\lambda_j) p_i(\lambda_j) = 1/4, \forall i \\
  &\sum_i p_i(\lambda_j) = 1, \forall j, \\
  &\sum_j q(\lambda_j) = 1.\\
\end{aligned}
\end{equation}

In the following, we investigate the optimal strategy based on value of $P$.

(1) when $\frac{1}{4}\le P\le \frac{1}{3}$.

With the normalization condition of $p_i(\lambda_j)$, we can rewrite $J$ as
\begin{equation}\label{eq:Proof4}
\begin{aligned}
 J_0 =& -\frac{1}{8} + \frac{1}{2}[q(\lambda_3)p_0(\lambda_3) + q(\lambda_4)p_0(\lambda_4) + q(\lambda_5)p_0(|\lambda_5)],\\
  J_1 =& -\frac{1}{8} +\frac{1}{2}[q(\lambda_1)p_1(\lambda_1) +2q(\lambda_2)p_1(\lambda_2) +2q(\lambda_3)p_1(\lambda_3)\\
  &+2q(\lambda_5)p_2(\lambda_5)],\\
  J_2= & -\frac{1}{8} +\frac{1}{2}[2q(\lambda_1)p_2(\lambda_1) +q(\lambda_2)p_2(\lambda_2) +2q(\lambda_4)p_2(\lambda_4)\\
  &+ 2q(\lambda_5)p_2(\lambda_5)],\\
  J_3 =& -\frac{1}{4} +\frac{1}{2}[2q(\lambda_1)p_3(\lambda_1) +2q(\lambda_2)p_3(\lambda_2) +2q(\lambda_3)p_3(\lambda_3)\\
  &+2q(\lambda_4)p_3(\lambda_4) ].\\
\end{aligned}
\end{equation}

In this case, we can write $J$ by
\begin{equation}\label{Eq:appendixbeta}
  J = 4\left(-\frac{5}{8} + \sum_{ij} \beta_{ij} q(\lambda_j)p_i(\lambda_j)\right),
\end{equation}
where the coefficient is given in Table~\ref{Table:coe2}.
\begin{table}[hbt]
\centering
\caption{The coefficient of $p_i(\lambda_j)$ in the expression of J.}
\begin{tabular}{cccccc}
  \hline
   &$q(\lambda_1)$&$q(\lambda_2)$&$q(\lambda_3)$&$q(\lambda_4)$&$q(\lambda_5)$\\
   \hline
   $p_0$&$0$&$0$&$\frac{1}{2}$&$\frac{1}{2}$&$\frac{1}{2}$\\
   $p_1$&$\frac{1}{2}$&$1$&$1$&$0$&$1$\\
   $p_2$&$1$&$\frac{1}{2}$&$0$&$1$&$1$\\
   $p_3$&$1$&$1$&$1$&$1$&$0$\\
  \hline
\end{tabular}\label{Table:coe2}
\end{table}

Note that $p_i$ is upper bounded by $P$, then we have
\begin{equation}\label{eq:Proof5}
\begin{aligned}
  J_0 &\leq -\frac{1}{8} + \frac{P}{2}\left[q(\lambda_3)+ q(\lambda_4) + q(\lambda_5)\right],\\
  J_1 &\leq -\frac{1}{8} +\frac{P}{2}[q(\lambda_1) +2q(\lambda_2) +2q(\lambda_3)+ 2q(\lambda_5)],\\
  J_2 &\leq -\frac{1}{8} +\frac{P}{2}[2q(\lambda_1)+q(\lambda_2) +2q(\lambda_4)+ 2q(\lambda_5)],\\
  J_3 &\leq -\frac{1}{4} +\frac{P}{2}[2q(\lambda_1)+2q(\lambda_2) +2q(\lambda_3)+2q(\lambda_4)].\\
\end{aligned}
\end{equation}
Therefore, we have
\begin{equation}\label{eq:Proof6}
\begin{aligned}
  J &\leq -4\{\frac{5}{8} + \frac{5P}{2}[q(\lambda_1)+q(\lambda_2) +q(\lambda_3)+q(\lambda_4)+q(\lambda_5)]\}\\
  &= \frac{5}{2}(4P-1).
\end{aligned}
\end{equation}

In addition, we can see that the equality holds by simply letting $p_i(\lambda_j)$ to be $P$ for $\beta_{i,j}\neq0$ and $p_i(\lambda_j)$ to be $1-3P$ for $\beta_{i,j}=0$. This special strategy is valid when $P\leq 1/3$, we have to consider differently for the other cases.


(2) When $\frac{1}{3}\le P\le \frac{3}{8}$.

With the constraints defined in Eq.~\eqref{eq:Proof4}, we can also write $J$ as follows,
\begin{equation}\label{eq:Proof q1}
\begin{aligned}
J=&4\{-\frac{5}{8}+q(\lambda_1)[1-\frac{1}{2} p_1(\lambda_1)-p_0(\lambda_1)]\\
&+q(\lambda_2)[1-\frac{1}{2} p_2(\lambda_2)-p_0(\lambda_2)]\\
&+q(\lambda_3)[1-\frac{1}{2} p_0(\lambda_3)-p_2(\lambda_3)]\\
&+q(\lambda_4)[1-\frac{1}{2} p_0(\lambda_4)-p_1(\lambda_4)]\\
&+q(\lambda_5)[1-\frac{1}{2} p_0(\lambda_5)-p_3(\lambda_3)]\}.\\
\end{aligned}
\end{equation}
Then $J$ can be similarly expressed by
\begin{equation}\label{Eq:appendixbeta2}
  J = 4\left(\frac{3}{8} + \sum_{ij} \beta_{ij} q(\lambda_j)p_i(\lambda_j)\right),
\end{equation}
with coefficient defined in Table~\label{table:2}.

\begin{table}[hbt]
\centering
\caption{The coefficient of $p_i(\lambda_j)$ in the expression of J.}
\begin{tabular}{cccccc}
  \hline
   &$q(\lambda_1)$&$q(\lambda_2)$&$q(\lambda_3)$&$q(\lambda_4)$&$q(\lambda_5)$\\
   \hline
   $p_0$&$-1$&$-1$&$-\frac{1}{2}$&$-\frac{1}{2}$&$-\frac{1}{2}$\\
   $p_1$&$-\frac{1}{2}$&$0$&$0$&$-1$&$0$\\
   $p_2$&$0$&$-\frac{1}{2}$&$-1$&$0$&$0$\\
   $p_3$&$0$&$0$&$0$&$0$&$-1$\\
  \hline
\end{tabular}
\label{table:2}
\end{table}

The intuition to maximize Eq.~\eqref{Eq:appendixbeta2} is to assign smaller values to $p_i(\lambda_j)$ for smaller corresponding coefficients. Because $\frac{1}{3}\le P\le \frac{3}{8}$, we can see that
\begin{equation}\label{}
  \frac{1}{2} p_i(\lambda_j)+p_{i'}(\lambda_j)\ge \frac{1}{2}(1-2P), \forall i,j,j'.
\end{equation}
Therefore, the Bell value defined in Eq.~\eqref{Eq:appendixbeta2} can be upper bounded by
\begin{equation}\label{eq:Proof qi4}
\begin{aligned}
J&\le4\left[ -\frac{5}{8}+ \sum_i q(\lambda_i)\left(1-\frac{1-2P}{2}\right)\right],\\
&=4P-\frac{1}{2}.
\end{aligned}
\end{equation}
This equal sign can be achieved by following parameter:
\begin{equation}\label{eq:Proof qi5}
\begin{aligned}
&q(\lambda_1)=q(\lambda_2)=\frac{1}{2}-\frac{1}{8(1-2P)};\\
&q(\lambda_3)=q(\lambda_4)=\frac{1}{8(1-2P)}+\frac{1}{8P}-\frac{1}{2};\\
&q(\lambda_5)=1-\frac{1}{4P};\\
&p_0(\lambda_1)=p_0(\lambda_2)=0,p_0(\lambda_3)=p_0(\lambda_4)=p_0(\lambda_5)=1-2P;\\
&p_1(\lambda_1)=1-2P,p_1(\lambda_2)=p_1(\lambda_3)=p_1(\lambda_5)=P,p_1(\lambda_4)=0;\\
&p_2(\lambda_2)=1-2P,p_2(\lambda_1)=p_1(\lambda_4)=p_1(\lambda_5)=P,p_2(\lambda_3)=0;\\
&p_3(\lambda_1)=p_3(\lambda_2)=p_3(\lambda_4)=p_3(\lambda_4)=p,p_3(\lambda_5)=0;\\
\end{aligned}
\end{equation}

(3) When $P\ge \frac{3}{8}$.

For this case, we can easily see that maximal Bell value can be achieved to be $1$, which is the algebra maximum of $J$. We show in the following that the Bell value cannot exceed $1$.

From Eq.~\eqref{Eq:appendixbeta2}, we know that $J$ can be expressed by
\begin{equation}\label{eq:Proof qi1}
\begin{aligned}
J=&\frac{3}{2}-4N\\
\end{aligned}
\end{equation}
where $N$ denotes the part contribute negatively,
\begin{equation}\label{eq:Proof qi2}
\begin{aligned}
N
&=\frac{1}{2}\sum_i q(\lambda_i) p_0(\lambda_i)+\frac{1}{2}[q(\lambda_1)p_0(\lambda_1)+q(\lambda_1) p_1(\lambda_1)\\
&+q(\lambda_2)p_0(\lambda_2)+q(\lambda_2) p_2(\lambda_2)+2q(\lambda_3)p_2(\lambda_3)\\
&+2q(\lambda_4)p_1(\lambda_4)+q(\lambda_5)p_3(\lambda_5)]\\
&=\frac{1}{8}+\frac{1}{2}[q(\lambda_1)(p_0(\lambda_1)+p_1(\lambda_1))+q(\lambda_2)(p_0(\lambda_2)+p_2(\lambda_2))\\
&+2q(\lambda_3)p_2(\lambda_3)+2q(\lambda_4)p_1(\lambda_4)+q(\lambda_5)p_3(\lambda_5)]\\
&\ge \frac{1}{8}.
\end{aligned}
\end{equation}
Therefore, we show that $J\leq1$. The equal sign is satisfied with the following strategy
\begin{equation}\label{eq:Proof qi3}
\begin{aligned}
&q(\lambda_1)=q(\lambda_2)=0;\\
&p_0(\lambda_3)=p_0(\lambda_4)=p_0(\lambda_5)=\frac{1}{4};\\
&p_1(\lambda_3)=p_1(\lambda_5)=\frac{3}{8},p_1(\lambda_4)=0;\\
&p_2(\lambda_4)=p_2(\lambda_5)=\frac{3}{8},p_2(\lambda_3)=0;\\
&p_3(\lambda_3)=p_1(\lambda_4)=\frac{3}{8},p_3(\lambda_5)=0.\\
\end{aligned}
\end{equation}

\subsection{$Q\neq0$}\label{P,Q}
In this part, we consider the input randomness quantification of $p_i(\lambda_j)$ with both $P$ and $Q$, which are defined in Eq.~\eqref{eq:randomness}. In this case, we have
\begin{equation}\label{}
  Q\leq p_i(\lambda_j) \leq P, \forall x,y,\lambda
\end{equation}

Note that, if we we substitute $p_i(\lambda_j)$ by
\begin{equation}\label{Eq:apppp}
  p'_i(\lambda_j)=\frac{p_i(\lambda_j)-Q}{1-4Q},
\end{equation}
we can show that the constraints on $p'_i(\lambda_j)$ are given by
 \begin{equation}\label{eq:constraints qi}
\begin{aligned}
  &0\le p'_i(\lambda_j)\le P' = \frac{P-Q}{1-4Q},\forall i,j \\
  &\sum_i q(\lambda_i) p'_j(\lambda_i) = 1/4, \forall j\\
  &\sum_i p'_i(\lambda_j) = 1, \forall j.
\end{aligned}
\end{equation}
Compared to Eq.\ref{Eq:appJ}, if we replace $p_i(\lambda_j)$ by $p'_i(\lambda_j)$, we obtain a new Bell value $J'$,
\begin{equation}\label{eq:def}
\begin{aligned}
  &J'/4\\
  &= q(\lambda_1)(p'_2(\lambda_1)-p'_0(\lambda_1))/2 + q(\lambda_2)(p'_1(\lambda_2)-p'_0(\lambda_2))/2\\
  &+q(\lambda_3)(p'_1(\lambda_3)-p'_2(\lambda_3))/2\\
  &+q(\lambda_4)(p'_2(\lambda_4)-p'_1(\lambda_4))/2+q(\lambda_5)[(p'_2(\lambda_5)+p'_1(\lambda_5))/2\\
  &-p'_3(\lambda_5)]
\end{aligned}
\end{equation}
Because $p_i(\lambda_j)$ and $p'_i(\lambda_j)$ are related by Eq.~\eqref{Eq:apppp}, we can prove that
\begin{equation}\label{Eq:appJJ}
  J(p_i(\lambda_j)) = (1-4Q)J'(p'_i(\lambda_j)).
\end{equation}

Therefore, instead of considering both upper and lower bound of $p_i(\lambda_j)$ in the original Bell's inequality, we can equivalently consider the same Bell inequality with $p'_i(\lambda_j)$, which has upper bound $P$ and lower bound $0$. We have our result as follows,

(1) When $\frac{P-Q}{1-4Q}\le \frac{1}{3}$, that is $3P+Q\le 1$
\begin{equation}
\begin{aligned}
J(P,Q)&=(1-4Q)J\left(\frac{P-Q}{1-4Q},0\right)\\
 &=(1-4Q)\frac{5}{2}\left(\frac{4P-4Q}{1-4Q}-1\right)\\
 &= \frac{5}{2}(4P-1)
 \end{aligned}
\end{equation}

(2) When $\frac{1}{3}\le \frac{P-Q}{1-4Q}\le \frac{3}{8} $, that is $3P+Q\ge 1$ and $2P+Q\le \frac{3}{4}$
\begin{equation}
\begin{aligned}
J(P,Q)&=(1-4Q)J\left(\frac{P-Q}{1-4Q}, 0\right)\\
 &=(1-4Q)\left(4\frac{P-Q}{1-4Q}-\frac{1}{2}\right)\\
 &= 4P-2Q-\frac{1}{2}\\
 \end{aligned}
\end{equation}

(3) When $ \frac{P-Q}{1-4Q}\ge \frac{3}{8} $, that is $2P+Q\ge \frac{3}{4}$
\begin{equation}
\begin{aligned}
J(P,Q)&=(1-4Q)J\left(\frac{P-Q}{1-4Q}, 0\right)\\
 &= 1-4Q\\
 \end{aligned}
\end{equation}

Therefore, the optimal CH value $J^{\mathrm{LHVM}}_{\mathrm{CH}}$ with LHVMs,
\begin{equation}\label{}
  J^{\mathrm{LHVM}}_{\mathrm{CH}}(P,Q) =
  \left\{
  \begin{array}{cc}
    \frac{5}{2}(4P-1) & 3P+Q\le 1\\
    1-4Q & 2P+Q\ge \frac{3}{4}\\
     4P-2Q-\frac{1}{2} & \mathrm{else}
  \end{array}
  \right.
\end{equation}

\section{Optimal strategy of the CH test: Factorizable condition}\label{app:fac}
Now, we consider the optimal strategy of the CH test with LHVMs under factorizable condition,
\begin{equation}\label{}
  p(i,j) = p_A(i)p_B(j).
\end{equation}
As we denote $p(i, j)$ by $p_{2*i+j}$, we have
 \begin{equation}\label{eq:Proof2}
\begin{aligned}
  p_0 & = p_A(0)p_B(0)\\
  p_1 & = p_A(0)p_B(1)\\
  p_2 & = p_A(1)p_B(0)\\
  p_3 & = p_A(1)p_B(1)\\
\end{aligned}
\end{equation}

\subsection{$Q=0$}
Similarly, we consider first the case with $Q=0$. In the following, we show that all the five possible strategies are upper bounded by $P-1/4$.

(1) When $P\le \frac{1}{2}$.

The result is based on the order of $p_1$, $p_2$, $p_3$, and $p_4$.

(a) $p_3 \geq p_2 \geq p_1 \geq p_0$ and $p_3 \geq p_1 \geq p_2 \geq p_0$.

This case is equivalent to $p_A(1) \geq p_A(0)$ and $p_B(1) \geq p_B(0)$. Thus we have $p_A(1)p_B(1) \leq P$. Amongst the five strategies, the biggest one is $ (p_2 - p_0)/2$, which can be upper bounded by
 \begin{equation}\label{eq:}
\begin{aligned}
  (p_2 - p_0)/2 & = (2p_A(1) - 1)(1 - p_B(1))/2,\\
   & \leq \frac{1}{2}[2p_A(1) + p_B(1) - 2p_A(1)p_B(1) - 1],\\
   & \leq P -\frac{1}{4}.
\end{aligned}
\end{equation}

(b) $p_1 \geq p_0 \geq p_3 \geq p_2$ and $p_3 \geq p_1 \geq p_2 \geq p_0$.

This case is equivalent to $p_A(0) \geq p_A(1)$ and $p_B(1) \geq p_B(0)$. Thus we have $p_A(0)p_B(1) \leq P$. Amongst the five strategies, the biggest one is  $(p_1 - p_2)/2$, which can be upper bounded by
 \begin{equation}\label{eq:}
\begin{aligned}
  (p_1 - p_2)/2 & = (p_A(0)p_B(1) -(1 - p_A(0))(1 - p_B(1)))/2,\\
   & \leq \frac{1}{2}[p_A(0) + p_B(1) - 1 ],\\
   & \leq P -\frac{1}{4}.
\end{aligned}
\end{equation}

(c) $p_2 \geq p_3 \geq p_0 \geq p_1$ and $p_2 \geq p_0 \geq p_3 \geq p_1$.

This case is equivalent to $p_A(1) \geq p_A(0)$ and $p_B(0) \geq p_B(1)$. Thus we have $p_A(1)p_B(0) \leq P$. Amongst the five strategies, the biggest one is  $(p_2 - p_1)/2$, which can be upper bounded by
 \begin{equation}\label{eq:}
\begin{aligned}
  (p_2 - p_1)/2 & =  (p_A(1)p_B(0) -(1 - p_A(1))(1 - p_B(0)))/2,\\
   & \leq \frac{1}{2}[p_A(1) + p_B(0) - 1 ],\\
   & \leq P -\frac{1}{4}.
\end{aligned}
\end{equation}

(d) $p_0 \geq p_1\geq p_2 \geq p_3$ and $p_0 \geq p_2 \geq p_1 \geq p_3$.

This case is equivalent to $p_A(0) \geq p_A(1)$ and $p_B(0) \geq p_B(1)$. Thus we have $p_A(0)p_B(0) \leq P$. Amongst the five strategies, the biggest one is  $(p_1 + p_2)/2 - p_3$, which can be upper bounded by
 \begin{equation}\label{eq:}
\begin{aligned}
  &(p_1 + p_2)/2 - p_3 \\
  & =p_A(0)(1 - p_B(0))/2 + (1 - p_A(0))p_B(0)/2 \\
  &- (1 - p_A(0))(1 -  p_B(0)),\\
   & \leq \frac{3}{2}[p_A(0) + p_B(0)] - 2p_A(0)p_B(0) - 1,\\
   & \leq P -\frac{1}{4}.
\end{aligned}
\end{equation}

Therefore, we show that all the strategies are upper bounded by $P-1/4$. Then the total Bell value
\begin{equation}\label{}
  J\leq 4(P-1/4) = 4P - 1,
\end{equation}
and the equal sign holds.

(2) When $P\ge \frac{1}{2}$.

It is easy to see that the maximal Bell value $J$ reaches 1 when $P\ge \frac{1}{2}$.

Consequently, we show the optimal Bell value  $J$ with LHVMs,
\begin{equation}\label{}
  J(P) =
  \left\{
  \begin{array}{cc}
    (4P-1) & P\le \frac{1}{2}\\
    1 & P> \frac{1}{2}\\
  \end{array}
  \right.
\end{equation}

\subsection{$Q\neq 0$}
We can follow a similar way in Appendix~\ref{P,Q} to take account of nonzero $Q$.

(1) When $\frac{P-Q}{1-4Q}\le \frac{1}{2}$, that is $P+Q\le \frac{1}{2}$
\begin{equation}
\begin{aligned}
J(P,Q)&=(1-4Q)J\left(\frac{P-Q}{1-4Q}, 0\right)\\
 &=(1-4Q)\left(4\frac{P-Q}{1-4Q}-1\right)\\
 &= 4P-1.
 \end{aligned}
\end{equation}

(2) When $\frac{P-Q}{1-4Q}> \frac{1}{2}$, that is $P+Q> \frac{1}{2}$
\begin{equation}
\begin{aligned}
J(P,Q)&=(1-4Q)J\left(\frac{P-Q}{1-4Q}, 0\right)\\
 &= 1-4Q\\
 \end{aligned}
\end{equation}

Thus, the Bell value $J^{\mathrm{LHVM, Fac}}_{\mathrm{CH}}$ with LHVMs under factorizable condition is,
\begin{equation}\label{}
  J^{\mathrm{LHVM, Fac}}_{\mathrm{CH}}(P,Q) =
  \left\{
  \begin{array}{cc}
    4P-1 & P+Q\le \frac{1}{2}\\
    1-4Q & P+Q> \frac{1}{2}\\
  \end{array}
  \right.
\end{equation}

\section{Equivalence between the CH and CHSH inequalities when assuming NS}\label{app:equiva}
In this section, we prove that the CH and CHSH inequality are equivalent when NS is assumed. We refer to \cite{Rosset14} for detail discussion about the connection between CH and CHSH.

\begin{proof}
According to the inputs, we can divide the CHSH inequality into four parts. When inputs are $ij$, define :
\begin{equation}\label{}
  J_{ij} = \sum_{a,b\in\{0,1\}}(-1)^{ a + b + ij}p(a, b|i, j).
\end{equation}
Owing to the NS condition, $J_{ij}$ can be rewritten by probabilities with output $0$,
\begin{equation}\label{}
\begin{aligned}
  J_{ij} &= \sum_{a,b\in\{0,1\}}(-1)^{ a + b + ij}p(a, b|i, j),\\
          &=  (-1)^{ij}(p(0,0|i,j)-p(0,1|i,j)-p(1,0|i,j)+p(1,1|0i,j)),  \\
          &=   (-1)^{ij}(p(0,0|i,j)-(p_A(0|i)- p(0,0|i,j))-(p_B£¨0|j)\\
          &- p(0,0|i,j))+(p_A(1|i)-p_B(0|j)+p(0,0|i,j))),\\
          &=(-1)^{ij}(1+4p(0,0|i,j)-2p_A(0|i)-2p_B(0|j)).\\
\end{aligned}
\end{equation}

Therefore, we have
\begin{equation}\label{}
\begin{aligned}
J_{\mathrm{CHSH}}&= \sum_{ij} J_{ij},\\
&=\sum_{ij} (-1)^{ij}(1+4p(0,0|i,j)-2p_A(0|i)-2p_B(0|j)),\\
 &= 2+4(p(0,0|0,0)+p(0,0|0,1)+p(0,0|1,0),\\
 &-p(0,0|1,1)-p_A(0|0)-p_B(0|0)),\\
 &= 2+4J_{\mathrm{CH}}.
 \end{aligned}
\end{equation}

\end{proof}
Hence, under the NS assumption, the value of the CH and the CHSH inequality are linearly related.  To analyze the best LHVMs strategy for the CH test, we can therefore consider the CHSH Bell test instead.

\section{Optimal strategy for the CHSH inequality}\label{app:CHSH}
Follwing the similar method described above, we first consider deterministic strategies, i.e., $p_A(0|x),p_B(0|y)\in\{0,1\}$ for the reason that any probabilistic LHVM could be realized with convex combination of deterministic ones.
Denote $p(i,j)$ as $p_{2*i + j}$, it is easy to show that the possible optimal deterministic strategies for $J_\lambda$ are
\begin{equation}
\begin{aligned}
   \{p_0+p_1+p_2-p_3&, p_0+p_1+p_3-p_2, \\
   p_0+p_2+p_3-p_1&, p_1+p_2+p_3-p_0\},
\end{aligned}
\end{equation}

\subsection{$Q=0$}
Here, we also first consider that $Q=0$.

(1) When $P\le \frac{1}{3}$.

We can show that, all the four strategies are upper bounded by $6P-1$. Take  the strategy of $p_0+p_1+p_2-p_3$ as an example,

\begin{equation}
\begin{aligned}
   p_0+p_1+p_2-p_3&\le P+P+P-(1-3P)\\
   &=6P-1.
\end{aligned}
\end{equation}

In this case, we can see that the CHSH value $J$ is upper bounded by $4(6P-1)$.

(2) When $P> \frac{1}{3}$.

In this case, LHVMs reaches the maximum Bell value, that is  $J$ can be 4.

Thus, the Bell value $J$ LHVMs is
\begin{equation}\label{}
  J(P) =
  \left\{
  \begin{array}{cc}
    24P-4 & P\le \frac{1}{3},\\
    4 & P> \frac{1}{3}.
  \end{array}
  \right.
\end{equation}

\subsection{$Q\neq0$}
For the case that $Q$ is nonzero, we apply the same transformation as Appendix~\ref{P,Q}. After the transformation defined in Eq.~\eqref{Eq:apppp}, the relation between $J(p_i(\lambda_j))$ and $J'(p'_i(\lambda_j))$ is given by
\begin{equation}\label{Eq:appJJ2}
  J(p_i(\lambda_j)) = (1-4Q)J'(p'_i(\lambda_j)) + 8Q.
\end{equation}
In this case, the optimal Bell value $J^{\mathrm{LHVM}}_{\mathrm{CHSH}}$ for the CHSH inequality with LHVMs is
\begin{equation}\label{}
  J^{\mathrm{LHVM}}_{\mathrm{CHSH}}(P,Q) =
  \left\{
  \begin{array}{cc}
    24P-4 & 3P+Q\le 1,\\
    4-8Q & 3P+Q\ge 1.
  \end{array}
  \right.
\end{equation}
And the optimal CH value $J^{\mathrm{LHVM, NS}}_{\mathrm{CH}}$ with LHVMs under NS is
\begin{equation}\label{}
  J^{\mathrm{LHVM, \mathrm{NS}}}_{\mathrm{CH}}(P,Q) =
  \left\{
  \begin{array}{cc}
    6P-3/2 & 3P+Q\le 1,\\
    1/2-2Q & 3P+Q\ge 1.
  \end{array}
  \right.
\end{equation}

\section{Optimal strategy for the CHSH inequality with factorizable condition}\label{app:NSFac}
In addition, we consider the factorizable condition,
\begin{equation}\label{}
  p(i,j) = p_A(i)p_B(j).
\end{equation}
In this case, we have
 \begin{equation}\label{eq:}
\begin{aligned}
  p_0 & = p_A(0)p_B(0)\\
  p_1 & = p_A(0)p_B(1)\\
  p_2 & = p_A(1)p_B(0)\\
  p_3 & = p_A(1)p_B(1)\\
\end{aligned}
\end{equation}

\subsection{$Q=0$}

(1) When $P\leq\frac{1}{2}$.

For the case that $Q=0$, $p_i$ are upper bounded by $P$ only. As the four strategies are symmetric, suppose that $p_3$ is the smallest one, which  is equivalent to $p_A(0) \geq p_A(1)$ and $p_B(0) \geq p_B(1)$. Thus we can see that $p_0+p_1+p_2-p_3$ is the largest strategy and is also upper bounded by
 \begin{equation}\label{eq:Proof2}
\begin{aligned}
  p_0+p_1+p_2-p_3 & = p_A(0) +(1 - p_A(0))(2p_B(0)-1)\\
   &=1-2(1-p_A(0))(1 - p_B(0),\\
   & \leq 1-(1-2P),\\
   &=2P
\end{aligned}
\end{equation}

Thus, we see that all the strategies are upper bounded by $2P$. Then the Bell value is upper bounded by $8P$.

(2) When$P> \frac{1}{2}$.

We can easily see that LHVMs reaches the maximum Bell value, that is  $J$ can be 4.

Consequently, the CHSH Bell value $J$ with LHVMs with factorizable condition is given by,
\begin{equation}\label{}
  J^{\mathrm{LHVM}}_{\mathrm{CHSH}}(P) =
  \left\{
  \begin{array}{cc}
    8P & P\le \frac{1}{2}\\
    4 & P> \frac{1}{2}\\
  \end{array}
  \right.
\end{equation}

\subsection{$Q\neq0$}
For the case where $Q\neq0$, we can similarly derive our result. The  CHSH Bell value $J^{\mathrm{LHVM,Fac}}_{\mathrm{CHSH}}(P,Q)$ is given by
\begin{equation}\label{}
  J^{\mathrm{LHVM,Fac}}_{\mathrm{CHSH}}(P,Q) =
  \left\{
  \begin{array}{cc}
    8P & P+Q \le \frac{1}{2},\\
    4-8Q & P+Q> \frac{1}{2}.\\
  \end{array}
  \right.
\end{equation}
And the optimal CH value $J^{\mathrm{LHVM, NS, Fac}}_{\mathrm{CH}}$ with LHVMs under NS and factorizable condition is
\begin{equation}\label{}
  J^{\mathrm{LHVM, NS, Fac}}_{\mathrm{CH}}(P,Q) =
  \left\{
  \begin{array}{cc}
    2P-1/2 & P+Q \le \frac{1}{2},\\
    1/2-2Q & P+Q> \frac{1}{2}.\\
  \end{array}
  \right.
\end{equation}

\bibliographystyle{apsrev4-1}

%


\end{document}